\documentclass[conference]{IEEEtran}
\IEEEoverridecommandlockouts

\usepackage{amsmath}
\usepackage{amssymb}
\usepackage{hyperref} 

\usepackage{url}  
\usepackage[capitalise]{cleveref}
\usepackage{graphicx}
\usepackage[utf8]{inputenc}
\usepackage{listings}
\usepackage[numbers]{natbib}
\usepackage{pgfplots}
\usepackage[binary-units]{siunitx}
\usepackage{subcaption}
\usepackage{tikz}
\usepgfplotslibrary{groupplots}
\usepackage{xcolor}

\DeclareMathOperator*{\argmin}{arg\,min}

\DeclareSIUnit\px{px}
\DeclareSIUnit\flop{Flop}
\DeclareSIUnit\process{process}
\crefname{lstlisting}{Listing}{Listings}
\Crefname{lstlisting}{Listing}{Listings}

\pgfplotsset{
    compat=newest,
    BaseAxis/.style={
        height=6cm,
        width=0.98\linewidth,
        xtick=data, 
        x label style={font=\scriptsize, yshift=0.3em},
        log basis x={2},
        log basis y={10},
        log origin=infty,
        y label style={font=\scriptsize, yshift=-0.3em},
        legend pos=north west,
        legend cell align=left,
        legend style={font=\scriptsize, inner sep=1pt},
        legend columns=2,
        grid=both,
        grid style={densely dotted, line width=0.3pt, draw=gray},
        every tick label/.append style={font=\scriptsize},
        xticklabel=\pgfmathparse{2^\tick}\pgfmathprintnumber{\pgfmathresult},
        legend columns=2,
        legend style={at={(0.5,-0.175)},anchor=north},
    },
    BasePlot/.style={
        line width=0.75pt,
        mark size=2.5pt,
        mark options={solid, draw=black, line width=0.75pt}, 
        error bars/.cd, y dir=both, y explicit,
        error bar style={line width=1pt,solid},
    },
    BaseGroupPlot/.style={
        height=6cm,
        width=0.48\linewidth,
        xtick=data, 
        x label style={font=\scriptsize, yshift=0.3em},
        log basis x={2},
        log basis y={10},
        log origin=infty,
        y label style={font=\scriptsize, yshift=-0.3em},
        grid=both,
        grid style={densely dotted, line width=0.3pt, draw=gray},
        every tick label/.append style={font=\scriptsize},
        xticklabel=\pgfmathparse{2^\tick}\pgfmathprintnumber{\pgfmathresult},
    }
}
\usetikzlibrary{
    arrows,
    calc,
    math,
    positioning,
    shapes,
}

\tikzset{
    bsp sync/.style={
        draw, black, fill=black, minimum height=6em, minimum width=0.2em, line width=0, inner sep=0
    },
    bsp proc line/.style={
        hgf-gray, densely dashed, ultra thin
    },
    bsp label/.style={
        font=\tiny\itshape
    },
    bsp compute/.style={
        draw, hgf-blue, fill=hgf-blue!35, minimum height=0.5em, ultra thin, inner sep=0, anchor=west
    },
    bsp communication/.style={
        hgf-blue, ->, >=stealth, shorten >=2pt, shorten <=2pt
    }
}

\definecolor{hgf-blue}{RGB}{0,90,160}
\definecolor{hgf-green}{RGB}{140,180,35}
\definecolor{hgf-gray}{RGB}{90,105,110}
\definecolor{hgf-purple}{RGB}{160,35,90}
\definecolor{hgf-red}{RGB}{210,50,100}
\definecolor{hgf-orange}{RGB}{240,120,30}
\definecolor{hgf-yellow}{RGB}{255,210,40}
\definecolor{hgf-turqoise}{RGB}{80,200,170}
\definecolor{hgf-dark-green}{RGB}{50,100,105}

\def\BibTeX{{\rm B\kern-.05em{\sc i\kern-.025em b}\kern-.08em
    T\kern-.1667em\lower.7ex\hbox{E}\kern-.125emX}}

\begin{document}
\bstctlcite{BSTcontrol}

\title{HeAT -- a Distributed and GPU-accelerated Tensor Framework for Data Analytics
\thanks{This work is supported by the Helmholtz Association Initiative and Networking Fund (INF) under project number ZT-I-0003 and under the Helmholtz AI platform grant.}
}

\author{\IEEEauthorblockN{
Markus Götz\IEEEauthorrefmark{4}, Charlotte Debus\IEEEauthorrefmark{1}, Daniel Coquelin\IEEEauthorrefmark{2}\IEEEauthorrefmark{3}\IEEEauthorrefmark{4}, Kai Krajsek\IEEEauthorrefmark{3}, Claudia Comito\IEEEauthorrefmark{3}, Philipp Knechtges\IEEEauthorrefmark{1}, \\
Björn Hagemeier\IEEEauthorrefmark{3}, Michael Tarnawa\IEEEauthorrefmark{3}, 
Simon Hanselmann\IEEEauthorrefmark{4},
Martin Siggel\IEEEauthorrefmark{1}, Achim Basermann\IEEEauthorrefmark{1} and Achim Streit\IEEEauthorrefmark{4}}

\IEEEauthorblockA{\IEEEauthorrefmark{1}\textit{Institute for Software Technology (SC)} \\
\textit{German Aerospace Center (DLR)}\\
Cologne, Germany \\
\{charlotte.debus, philipp.knechtges, martin.siggel, achim.basermann\}@dlr.de}
\IEEEauthorblockA{\IEEEauthorrefmark{2}\textit{Institute of Bio- and Geosciences Agrosphere (IBG-3)} \\
\textit{Forschungszentrum J\"{u}lich (FZJ)}\\
Jülich, Germany}
\IEEEauthorblockA{\IEEEauthorrefmark{3}\textit{J\"{u}lich Supercomputing Centre (JSC)} \\
\textit{Forschungszentrum J\"{u}lich (FZJ)}\\
Jülich, Germany \\
\{k.krajsek, c.comito, b.hagemeier, m.tarnawa\}@fz-juelich.de}
\IEEEauthorblockA{\IEEEauthorrefmark{4}\textit{Steinbuch Centre for Computing (SCC)} \\
\textit{Karlsruhe Institute of Technology (KIT)}\\
Karlsruhe, Germany \\
\{markus.goetz, daniel.coquelin, simon.hanselmann, achim.streit\}@kit.edu}
}


\maketitle


\begin{abstract}
To cope with the rapid growth in available data, the efficiency of data analysis and machine learning libraries has recently received increased attention. Although great advancements have been made in traditional array-based computations, most are limited by the resources available on a single computation node. Consequently, novel approaches must be made to exploit distributed resources, e.g. distributed memory architectures. To this end, we introduce HeAT, an array-based numerical programming framework for large-scale parallel processing with an easy-to-use NumPy-like API. HeAT utilizes PyTorch as a node-local eager execution engine and distributes the workload on arbitrarily large high-performance computing systems via MPI. It provides both low-level array computations, as well as assorted higher-level algorithms. With HeAT, it is possible for a NumPy user to take full advantage of their available resources, significantly lowering the barrier to distributed data analysis. When compared to similar frameworks, HeAT achieves speedups of up to two orders of magnitude.
\end{abstract}


\section{Introduction}
\label{sec:introduction}
In the age of Big Data, modern-day computational data science heavily relies on generating highly complex data-driven models. However, the consistent increase in data volume is challenging the processing power for data analytics and machine learning (ML) frameworks. 

The Python programming language has evolved into the de-facto standard for the data science community. Therein, the default choice for many applications is the SciPy stack~\cite{viranen2020scipy}, which is built upon the computational library NumPy~\cite{walt2011numpy}. More recently, deep-learning libraries such as TensorFlow~\cite{abadi2015tensorflow} and PyTorch~\cite{paszke2019pytorch}, have begun to bridge the gap between small scale workstation-based computing and high-performance multi-node computing by offering GPU-accelerated kernels. Although these frameworks include options for manually distributing computations, they are generally confined to the processing capabilities of a singular computation node. As the size of datasets increase, single-node computations are at best impractical and at worst impossible. 

In response to these challenges, we propose HeAT\footnote{\url{https://github.com/helmholtz-analytics/heat}} -- the Helmholtz Analytics Toolkit: an open-source library with a NumPy-like API for distributed and GPU-accelerated computing on general-purpose clusters and high performance computing (HPC) systems. HeAT implements parallel algorithms for both low-level array computations as well as higher-level data analytics and machine learning methods. It does so by interweaving process-local PyTorch tensors with communication via the Message Passing Interface (MPI)~\cite{mpi2015mpi}. The familiar API facilitates the transition of existing Python code to distributed applications, thus opening the doorway to HPC computing for domain-specialized scientists. Due to its design, HeAT consistently performs better in terms of execution time and scalability when compared to similar frameworks.

The remainder of this paper is organized as follows. \cref{sec:related-work} will present related work in the field of distributed and GPU-accelerated array computations in Python. \cref{sec:framework-design-and-implementation} will explain HeAT's programming model, as well as its array and communication design concepts. In \cref{sec:performance-result}, an empirical performance evaluation is presented. \cref{sec:discussion} discusses the advantages and limitations of HeAT's programming model with respect to other frameworks. Finally, \cref{sec:conclusion} concludes the presented work.

\section{Related Work}
\label{sec:related-work}

NumPy is arguably the most widely used Python-based data science library. It offers powerful data structures and algorithms for vectorized matrix and tensor operations, allowing for the implementation of efficient numerical and scientific programs. The high popularity of NumPy, likely due to the similarity of its functions to the mathematical formulation, has led to its API representing a widely recognized and accepted standard for data science programming. 
Scikit-learn \cite{pedregosa2011sklearn} is a NumPy-based machine learning framework that offers a wide range of ready-to-use high-level algorithms for clustering, classification and regression, as well as selected pre-processing steps like feature selection and dimensionality reduction. This makes it a very attractive solution for application scientists. However, neither NumPy nor scikit-learn support out-of-the-box GPU usage.

Several packages have addressed GPU acceleration for array computations. The just-in-time compiler Numba \cite{lam2015numba} can compile Python functions to CUDA code. While Numba relies on custom annotation of the Python code with decorators for acceleration, the CUDA-accelerated array computation library CuPy \cite{nishino2017cupy} allows for GPU computation in analogy to the NumPy interface. Unfortunately, it does not provide the full range of NumPy functionality. RAPIDS's~\cite{rapids} CUDA-accelerated libraries CuDF and CuML aim towards higher-level machine learning functionality beyond low-level array computation. CuDF offers GPU dataframe computations similar to the Pandas library~\cite{reback2020pandas}. CuML offers high-level machine learning algorithms to some extent. However, the currently available function space of these libraries is limited.

With the increasing interest in deep learning methods, novel frameworks have emerged that place special focus onto tensor linear algebra and neural networks. Many libraries like PyTorch \cite{paszke2019pytorch}, TensorFlow~\cite{abadi2015tensorflow}, MXNet~\cite{chen2015mxnet} or JAX~\cite{bradburg2018jax} focus mainly on deep learning applications, and as such they do not actively target NumPy-API compatibility. On the other hand, they enable simple transitions between CPU and GPU computation. Unfortunately, they primarily provide high-level algorithms for neural networks and often lack many traditional algorithms, for example clustering or ensemble methods. 

All of the aforementioned libraries work around Python's parallel computation limitations by implementing computationally intensive kernels in low-level programming languages, like C++ or CUDA, and invoking the respective calls at run-time. This allows Python users to exploit powerful features such as vectorization, threading, or the utilization of accelerator hardware. However, these frameworks are designed for single computation node usage, with specific configurations enabling multi-core computation. This limits their potential application to small and medium size problems. For larger datasets, excessive computation times and increasing memory consumption pose arduous challenges.

Some of these ML libraries provide a basic infrastructure for distributed computation. PyTorch, for example, includes two such packages. Firstly, the distributed remote procedure call (RPC) framework handles communication and by that enables distributed automatic differentiation (AD). Secondly, the \texttt{distributed} package implements low-level bindings to MPI, Gloo, and NCCL. However the set of MPI functionality is not complete as it targets the communication functions specifically required for data-parallel neural networks. Furthermore, the communication aspect of algorithms must be implemented by the user. This requires, at minimum, a basic understanding of distributed computing. Similar restrictions apply for the distributed extensions of TensorFlow, JAX, and MXNet.
Frameworks like DeepSpeed~\cite{rajbh2019zero} and Horovod~\cite{sergeev2018horovod} offer multi-node computations on both CPU and GPU. These approaches are limited to deep learning applications, as they mainly focus on data-parallelism for neural networks and do not target general distributed array-based computations. 

Phylanx~\cite{tohid2018phylanx} implements Python bindings for C and C++ array computation algorithms, which closely mimic NumPy's API, but it does not support higher-level machine learning functionality or GPU usage. Intel’s DAAL~\cite{chen2017benchmarking} provides only high-level algorithms employing MapReduce~\cite{dean2008mapreduce}, with the focus on accelerated multi-CPU usage. However, it does not offer the means for low-level array computations or GPU usage. Furthermore, functionality requiring communication beyond the scope of MapReduce must be implemented by the user. Legate~\cite{bauer2019legate} focuses on distributed multi-node multi-GPU computations for low- to high-level array operations by employing a global address space programming interface. While its design enables shared-distributed memory computation, a wide range of functionality is currently not  implemented.

When it comes to easy distributed array computations with a NumPy-like API and support of high-level algorithms similar to scikit-learn, Dask~\cite{rocklin2015dask} has become the most popular framework amongst application scientists. It employs dynamic task scheduling for parallel execution of NumPy operations on CPU-based multi-node systems. The Dask execution model envisages one scheduler process and multiple worker instances. The scheduler sends workload via serialized RPC calls to the workers. Networking between the processes builds on the available system network architecture. GPU usage can be enabled by coupling Dask to RAPIDS's cuML and cuDF. Due to its popularity, Dask can be considered the current benchmark in Python-first distributed data analysis and ML computations. There are non-Python based distributed data analysis tools such as the Java-based Spark~\cite{spark}. A recent comparison~\cite{sparkvdask} between Spark and Dask has shown that they preform similarly. 

\section{Design and Implementation}
\label{sec:framework-design-and-implementation}

HeAT is an open-source library that implements data structures, functions, and methods for array-based numerical data analysis and machine learning. Due to its NumPy-like API, users are generally familiar with the programming approach. The implementation of (distributed) higher-level algorithms adheres to the  scikit-learn API. This interface design makes the conversion of existing data analytics applications to distributed HeAT applications straightforward. An example can be seen in \cref{lst:stddev}. Furthermore, small-scale program prototypes can be developed, which can be transitioned transparently to HPC systems without major code or algorithmic changes. Distributed HeAT applications are typically faster, and their memory limitations are those of the entire system, rather than those of a single node. As a result, HeAT facilitates the algorithmic development and efficient deployment of large-scale data analytics and machine learning applications. 
\begin{lstlisting}[
    caption={Implementation of a function calculating the standard deviation of an array, demonstrating the API compatibility between NumPy and HeAT.},
    label={lst:stddev}
]
import numpy as np
import heat as ht
    
def np_stddev(a, axis=0):
    return np.sqrt((a - a.mean(axis)) ** 2)
    
def ht_stddev(a, axis=0):
    return ht.sqrt((a - a.mean(axis)) ** 2)
\end{lstlisting}

The central component of HeAT is the \texttt{DNDarray} data structure, an N-dimensional array transparently composed of computational objects on one or more processes. The process-local objects are PyTorch tensors, allowing HeAT functions to use both CPUs and GPUs. A detailed description of the \texttt{DNDarray} design will be given in \ref{subsec:dndarrays}.

For distributed memory computing, communication between processes is crucial. MPI controls communication among parallel processes on distributed memory systems via a set of message sending and receiving functions. Communication can take place between two processes, i.e. point-to-point communication, or within groups of processes in the MPI communicator, i.e. global communication. HeAT's MPI-based custom communications backend is described in  \cref{subsec:distributed-computation}.

\subsection{Programming Model}
\label{subsec:programming-model}

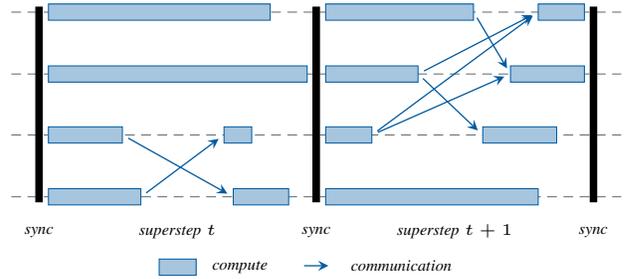
\begin{figure}[t]
    \resizebox{.95\linewidth}{!}{
    \centering
    \begin{tikzpicture}
    \draw[bsp proc line] (-0.3, 1.0) -- (6.3, 1.0);
    \draw[bsp proc line] (-0.3, 0.33) -- (6.3, 0.33);
    \draw[bsp proc line] (-0.3, -0.33) -- (6.3, -0.33);
    \draw[bsp proc line] (-0.3, -1.0) -- (6.3, -1.0);
    
    \node[bsp sync] (sync1) at (0.0, 0.0) {};
    \node[bsp sync] (sync2) at (3.0, 0.0) {};
    \node[bsp sync] (sync3) at (6.0, 0.0) {};
    
    \node[bsp label, below=0.4em of sync1] (sync1 label) {sync};
    \node[bsp label, below=0.4em of sync2] (sync2 label) {sync};
    \node[bsp label, below=0.4em of sync3] (sync3 label) {sync};
    
    \node[bsp label] (sst) at ($(sync1 label)!0.5!(sync2 label)$) {superstep $t$};
    \node[bsp label] (sstp1) at ($(sync2 label)!0.5!(sync3 label)$) {superstep $t+1$};
    
    \node[bsp compute, minimum width=2.4cm] (compute 1-1-1) at (0.1, 1.0) {};
    \node[bsp compute, minimum width=2.8cm] (compute 2-1-1) at (0.1, 0.33) {};
    \node[bsp compute, minimum width=0.8cm] (compute 3-1-1) at (0.1, -0.33) {};
    \node[bsp compute, minimum width=0.3cm] (compute 3-1-2) at (2.0, -0.33) {};
    \node[bsp compute, minimum width=1.0cm] (compute 4-1-1) at (0.1, -1.0) {};
    \node[bsp compute, minimum width=0.6cm] (compute 4-1-2) at (2.1, -1.0) {};
    
    \node[bsp compute, minimum width=1.6cm] (compute 1-2-1) at (3.1, 1.0) {};
    \node[bsp compute, minimum width=0.5cm] (compute 1-2-2) at (5.4, 1.0) {};
    \node[bsp compute, minimum width=1.0cm] (compute 2-2-1) at (3.1, 0.33) {};
    \node[bsp compute, minimum width=0.8cm] (compute 2-2-2) at (5.1, 0.33) {};
    \node[bsp compute, minimum width=0.5cm] (compute 3-2-1) at (3.1, -0.33) {};
    \node[bsp compute, minimum width=0.8cm] (compute 3-2-2) at (4.8, -0.33) {};
    \node[bsp compute, minimum width=2.3cm] (compute 4-2-1) at (3.1, -1.0) {};
    
    \draw[bsp communication] (compute 3-1-1.east) -- (compute 4-1-2.west);
    \draw[bsp communication] (compute 4-1-1.east) -- (compute 3-1-2.west);
    
    \draw[bsp communication] (compute 1-2-1.east) -- (compute 2-2-2.west);
    \draw[bsp communication] (compute 2-2-1.east) -- (compute 1-2-2.west);
    \draw[bsp communication] (compute 2-2-1.east) -- (compute 3-2-2.west);
    \draw[bsp communication] (compute 3-2-1.east) -- (compute 2-2-2.west);
    \draw[bsp communication] (compute 3-2-1.east) -- (compute 1-2-2.west);
    
    \node[bsp compute, minimum width=0.4cm, below=0.1 of sst] (compute legend) {};
    \node[bsp label, right=0.05 of compute legend] {compute};
    
    \node[bsp compute, minimum width=0.4cm, below=0.1 of sync2 label, opacity=0] (comm legend) {};
    \draw[bsp communication] (comm legend.west) -- (comm legend.east);
    \node[bsp label, right=0.05 of comm legend] {communication};
\end{tikzpicture}}
    \caption{The BSP-inspired parallel processing model utilized by HeAT. Computation steps are marked as light blue blocks, possible communication as blue arrows, and implicit or explicit synchronization points as vertical black bars.}
    \label{fig:bsp}
\end{figure}

HeAT realizes a single-program-multiple-data (SPMD) programming model~\cite{darema2001spmd} using PyTorch and MPI. Additionally, the framework's processing model is inspired by the bulk-synchronous parallel (BSP)~\cite{valiant1990bsp} model. In practice, 
computations proceed in a series of hierarchical supersteps, each consisting of a number of process-local computations and subsequent inter-process communications. In contrast to the classical BSP model, communicated data is available immediately, rather than after the next global synchronization. In HeAT, global synchronizations only occur for collective MPI calls as well as at the program start and termination. A schematic overview is depicted in \cref{fig:bsp}.

The process-local computations are implemented using PyTorch as the computation engine. Each computation is processed eagerly, i.e. when issued to the interpreter. The scheduling onto the hardware is controlled by the respective runtime environment of PyTorch. For the CPU backend, these are the synchronous schedulers of OpenMP~\cite{dagum1998openmp} and Intel TBB~\cite{pheatt2008tbb}. For the GPU backend, it is the asynchronous scheduler of the NVIDIA CUDA~\cite{nickolls2008cuda} runtime system. HeAT provides the MPI "glue", utilizing the \texttt{mpi4py}~\cite{dalcin08mpi4py} module, for the communication in each superstep. Users can freely access these implementation details, although it is neither necessary nor recommended to modify the communication routines.

\subsection{DNDarrays}
\label{subsec:dndarrays}

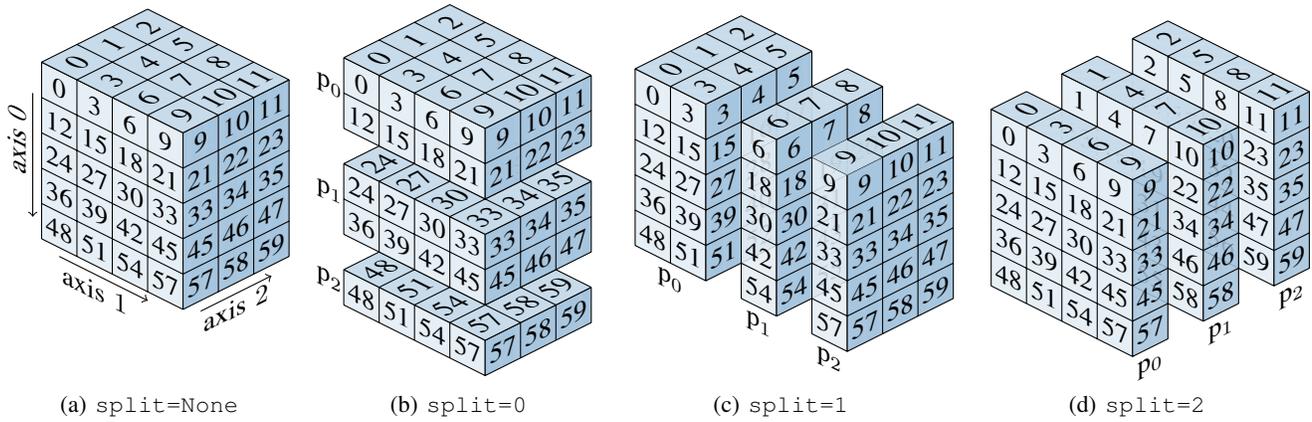
\begin{figure*}
    \begin{subfigure}[b]{0.22\linewidth}
        \begin{tikzpicture}[scale=0.47][
    every node/.style={minimum size=1cm},
    on grid
]
\begin{scope}[
    every node/.append style={yslant=-0.5},
    yslant=-0.5
]
    \shade[right color=hgf-blue!10, left color=hgf-blue!10] (0,0) rectangle +(4,5);
    \node at (0.5,4.5) {0};
    \node at (1.5,4.5) {3};
    \node at (2.5,4.5) {6};
    \node at (3.5,4.5) {9};
    \node at (0.5,3.5) {12};
    \node at (1.5,3.5) {15};
    \node at (2.5,3.5) {18};
    \node at (3.5,3.5) {21};
    \node at (0.5,2.5) {24};
    \node at (1.5,2.5) {27};
    \node at (2.5,2.5) {30};
    \node at (3.5,2.5) {33};
    \node at (0.5,1.5) {36};
    \node at (1.5,1.5) {39};
    \node at (2.5,1.5) {42};
    \node at (3.5,1.5) {45};
    \node at (0.5,0.5) {48};
    \node at (1.5,0.5) {51};
    \node at (2.5,0.5) {54};
    \node at (3.5,0.5) {57};
    \draw (0,0) grid (4,5);
   \draw[->, line width=.4pt] (-0.25,4.) -- (-0.25,0.5);
   \node[rotate=90] at (-0.7,2.5) {axis $0$};
    \draw[->, line width=.4pt] (0.5,-0.25) -- (3.,-0.25);
    \node at (1.5, -0.7) {axis $1$};
\end{scope}

\begin{scope}[
    every node/.append style={yslant=0.5},
    yslant=0.5
]
    \shade[right color=hgf-blue!35,left color=hgf-blue!35] (4,-4) rectangle +(3,5);
    \node at (4.5,0.5) {9};
    \node at (5.5,0.5) {10};
    \node at (6.5,0.5) {11};
    \node at (4.5,-0.5) {21};
    \node at (5.5,-0.5) {22};
    \node at (6.5,-0.5) {23};
    \node at (4.5,-1.5) {33};
    \node at (5.5,-1.5) {34};
    \node at (6.5,-1.5) {35};
    \node at (4.5,-2.5) {45};
    \node at (5.5,-2.5) {46};
    \node at (6.5,-2.5) {47};
    \node at (4.5,-3.5) {57};
    \node at (5.5,-3.5) {58};
    \node at (6.5,-3.5) {59};
    \draw (4,-4) grid (7,1);
    \draw[->, line width=.4pt] (4.5,-4.3 ) -- (6.5,-4.3);
    \node at (5.5,-4.7) {axis $2$};
\end{scope}

\begin{scope}[
    every node/.append style={yslant=0.5, xslant=-1},
    yslant=0.5,
    xslant=-1
]
    \shade[bottom color=hgf-blue!22, top color=hgf-blue!22] (8,5) rectangle +(-3,-4);
    \node at (5.5,4.5) {0};
    \node at (5.5,3.5) {3};
    \node at (5.5,2.5) {6};
    \node at (5.5,1.5) {9};
    \node at (6.5,4.5) {1};
    \node at (6.5,3.5) {4};
    \node at (6.5,2.5) {7};
    \node at (6.5,1.5) {10};
    \node at (7.5,4.5) {2};
    \node at (7.5,3.5) {5};
    \node at (7.5,2.5) {8};
    \node at (7.5,1.5) {11};
\draw (8,5) grid (5, 1);
\end{scope}

\end{tikzpicture}
        \caption{\texttt{split=None}}
    \end{subfigure}
    \begin{subfigure}[b]{0.22\linewidth}
        \begin{tikzpicture}[scale=0.47][
    every node/.style={minimum size=1cm}, 
    on grid
]
\begin{scope}[
    every node/.append style={yslant=-0.5},
    yslant=-0.5
]
    \shade[right color=hgf-blue!10, left color=hgf-blue!10] (0,0) rectangle +(4,1);
    \node at (0.5,0.5) {48};
    \node at (1.5,0.5) {51};
    \node at (2.5,0.5) {54};
    \node at (3.5,0.5) {57};
    \draw (0,0) grid (4,1);
    \node[] at (-0.4,0.5) {p$_2$};
\end{scope}

\begin{scope}[
    every node/.append style={yslant=0.5},
    yslant=0.5
]
    \shade[right color=hgf-blue!35,left color=hgf-blue!35] (4,-4) rectangle +(3,1);
    \node at (4.5,-3.5) {57};
    \node at (5.5,-3.5) {58};
    \node at (6.5,-3.5) {59};
    \draw (4,-4) grid (7,-3);
\end{scope}

\begin{scope}[
    every node/.append style={yslant=0.5,xslant=-1},
    yslant=0.5,
    xslant=-1
]
    \shade[bottom color=hgf-blue!22, top color=hgf-blue!22] (4,1) rectangle +(-3,-4);
    \node at (1.5,0.5) {48};
    \node at (1.5,-0.5) {51};
    \node at (1.5,-1.5) {54};
    \node at (1.5,-2.5) {57};
    \node at (2.5,0.5) {49};
    \node at (2.5,-0.5) {52};
    \node at (2.5,-1.5) {55};
    \node at (2.5,-2.5) {58};
    \node at (3.5,0.5) {50};
    \node at (3.5,-0.5) {53};
    \node at (3.5,-1.5) {56};
    \node at (3.5,-2.5) {59};
    \draw (4,1) grid (1, -3);
\end{scope}

\begin{scope}[
    every node/.append style={yslant=-0.5},yslant=-0.5
]
    \shade[right color=hgf-blue!10, left color=hgf-blue!10] (0,2) rectangle +(4,2);
    \node at (0.5,3.5) {24};
    \node at (1.5,3.5) {27};
    \node at (2.5,3.5) {30};
    \node at (3.5,3.5) {33};
    \node at (0.5,2.5) {36};
    \node at (1.5,2.5) {39};
    \node at (2.5,2.5) {42};
    \node at (3.5,2.5) {45};
    \draw (0,2) grid (4,4);
     \node[] at (-0.4,3) {p$_1$};

\end{scope}

\begin{scope}[every node/.append style={yslant=0.5},yslant=0.5]
    \shade[right color=hgf-blue!35,left color=hgf-blue!35] (4,-2) rectangle +(3,2);
    \node at (4.5,-0.5) {33};
    \node at (5.5,-0.5) {34};
    \node at (6.5,-0.5) {35};
    \node at (4.5,-1.5) {45};
    \node at (5.5,-1.5) {46};
    \node at (6.5,-1.5) {47};
    \draw (4,-2) grid (7,0);
\end{scope}

\begin{scope}[
    every node/.append style={yslant=0.5,xslant=-1},yslant=0.5,xslant=-1
  ]
    \shade[bottom color=hgf-blue!22, top color=hgf-blue!22] (7,4) rectangle +(-3,-4);
    \node at (4.5,3.5) {24};
    \node at (4.5,2.5) {27};
    \node at (4.5,1.5) {30};
    \node at (4.5,0.5) {33};
    \node at (5.5,3.5) {25};
    \node at (5.5,2.5) {28};
    \node at (5.5,1.5) {31};
    \node at (5.5,0.5) {34};
    \node at (6.5,3.5) {26};
    \node at (6.5,2.5) {29};
    \node at (6.5,1.5) {32};
    \node at (6.5,0.5) {35};
    \draw (7,4) grid (4,0);
\end{scope}

\begin{scope}[
    every node/.append style={yslant=-0.5},yslant=-0.5
]
    \shade[right color=hgf-blue!10, left color=hgf-blue!10] (0,5) rectangle +(4,2);
    \node at (0.5,6.5) {0};
    \node at (1.5,6.5) {3};
    \node at (2.5,6.5) {6};
    \node at (3.5,6.5) {9};
    \node at (0.5,5.5) {12};
    \node at (1.5,5.5) {15};
    \node at (2.5,5.5) {18};
    \node at (3.5,5.5) {21};
    \draw (0,5) grid (4,7);
        \node[] at (-0.4,6) {p$_0$};

\end{scope}

\begin{scope}[
    every node/.append style={yslant=0.5},yslant=0.5
]
    \shade[right color=hgf-blue!35,left color=hgf-blue!35] (4,1) rectangle +(3,2);
    \node at (4.5,2.5) {9};
    \node at (5.5,2.5) {10};
    \node at (6.5,2.5) {11};
    \node at (4.5,1.5) {21};
    \node at (5.5,1.5) {22};
    \node at (6.5,1.5) {23};
    \draw (4,1) grid (7,3);
\end{scope}

\begin{scope}[
    every node/.append style={yslant=0.5,xslant=-1},
    yslant=0.5,
    xslant=-1
]
    \shade[bottom color=hgf-blue!22, top color=hgf-blue!22] (10,7) rectangle +(-3,-4);
    \node at (7.5,6.5) {0};
    \node at (7.5,5.5) {3};
    \node at (7.5,4.5) {6};
    \node at (7.5,3.5) {9};
    \node at (8.5,6.5) {1};
    \node at (8.5,5.5) {4};
    \node at (8.5,4.5) {7};
    \node at (8.5,3.5) {10};
    \node at (9.5,6.5) {2};
    \node at (9.5,5.5) {5};
    \node at (9.5,4.5) {8};
    \node at (9.5,3.5) {11};
    \draw (10,7) grid (7, 3);
\end{scope}
\end{tikzpicture}
        \caption{\texttt{split=0}}
    \end{subfigure}
    \hspace{.05cm}
    \begin{subfigure}[b]{0.22\linewidth}
    \captionsetup{skip=-10pt}
        \begin{tikzpicture}[scale=0.47][
    every node/.style={minimum size=1cm},
    on grid

]
\tikzmath{\dx = -2; \dy =-2;} 
\begin{scope}[
    every node/.append style={yslant=-0.5},
    yslant=-0.5
]
    \shade[right color=hgf-blue!10, left color=hgf-blue!10] (0,-2-\dy) rectangle +(2,5);
    \node at (0.5,2.5-\dy) {0};
    \node at (1.5,2.5-\dy) {3};
    \node at (0.5,1.5-\dy) {12};
    \node at (1.5,1.5-\dy) {15};
    \node at (0.5,0.5-\dy) {24};
    \node at (1.5,0.5-\dy) {27};
    \node at (0.5,-0.5-\dy) {36};
    \node at (1.5,-0.5-\dy) {39};
    \node at (0.5,-1.5-\dy) {48};
    \node at (1.5,-1.5-\dy) {51};
    \draw (0,-2-\dy) grid (2,3-\dy);
   \node at (1, -2.5-\dy) {p$_0$};  
\end{scope}

\begin{scope}[
    every node/.append style={yslant=0.5},
    yslant=0.5
]
    \shade[right color=hgf-blue!35,left color=hgf-blue!35] (2,-4-\dy) rectangle +(3,5);
    \node at (2.5,0.5-\dy) {3};
    \node at (3.5,0.5-\dy) {4};
    \node at (4.5,0.5-\dy) {5};
    \node at (2.5,-0.5-\dy) {15};
    \node at (3.5,-0.5-\dy) {16};
    \node at (4.5,-0.5-\dy) {17};
    \node at (2.5,-1.5-\dy) {27};
    \node at (3.5,-1.5-\dy) {28};
    \node at (4.5,-1.5-\dy) {29};
    \node at (2.5,-2.5-\dy) {39};
    \node at (3.5,-2.5-\dy) {40};
    \node at (4.5,-2.5-\dy) {41};
    \node at (2.5,-3.5-\dy) {51};
    \node at (3.5,-3.5-\dy) {52};
    \node at (4.5,-3.5-\dy) {53};
    \draw (2,-4-\dy) grid (5,1-\dy);
\end{scope}

\begin{scope}[
    every node/.append style={yslant=0.5,xslant=-1},
    yslant=0.5,
    xslant=-1
]
    \shade[bottom color=hgf-blue!22, top color=hgf-blue!22] (6-\dx,3-\dy) rectangle +(-3,-2);
    \node at (3.5-\dx,2.5-\dy) {0};
    \node at (3.5-\dx,1.5-\dy) {3};
    \node at (4.5-\dx,2.5-\dy) {1};
    \node at (4.5-\dx,1.5-\dy) {4};
    \node at (5.5-\dx,2.5-\dy) {2};
    \node at (5.5-\dx,1.5-\dy) {5};
    \draw (6-\dx,3-\dy) grid (3-\dx, 1-\dy);
\end{scope}

\begin{scope}[
    every node/.append style={yslant=-0.5},
    yslant=-0.5
]
    \shade[right color=hgf-blue!10, left color=hgf-blue!10] (3,-2-\dy) rectangle +(1,5);
    \node at (3.5,2.5-\dy) {6};
    \node at (3.5,1.5-\dy) {18};
    \node at (3.5,0.5-\dy) {30};
    \node at (3.5,-0.5-\dy) {42};
    \node at (3.5,-1.5-\dy) {54};
    \draw (3,-2-\dy) grid (4,3-\dy);
     \node at (3.5, -2.5-\dy) {p$_1$}; 
\end{scope}

\begin{scope}[
    every node/.append style={yslant=0.5},
    yslant=0.5
]
    \shade[right color=hgf-blue!35,left color=hgf-blue!35] (4,-6-\dy) rectangle +(3,5);
    \node at (4.5,-1.5-\dy) {6};
    \node at (5.5,-1.5-\dy) {7};
    \node at (6.5,-1.5-\dy) {8};
    \node at (4.5,-2.5-\dy) {18};
    \node at (5.5,-2.5-\dy) {19};
    \node at (6.5,-2.5-\dy) {20};
    \node at (4.5,-3.5-\dy) {30};
    \node at (5.5,-3.5-\dy) {31};
    \node at (6.5,-3.5-\dy) {32};
    \node at (4.5,-4.5-\dy) {42};
    \node at (5.5,-4.5-\dy) {43};
    \node at (6.5,-4.5-\dy) {44};
    \node at (4.5,-5.5-\dy) {54};
    \node at (5.5,-5.5-\dy) {55};
    \node at (6.5,-5.5-\dy) {56};
    \draw (4,-6-\dy) grid (7,-1-\dy);
\end{scope}

\begin{scope}[
    every node/.append style={yslant=0.5,xslant=-1},
    yslant=0.5,
    xslant=-1
]
    \shade[bottom color=hgf-blue!22, top color=hgf-blue!22] (6-\dx,0-\dy) rectangle +(-3,-1);
    \node at (3.5-\dx,-0.5-\dy) {6};
    \node at (4.5-\dx,-0.5-\dy) {7};
    \node at (5.5-\dx,-0.5-\dy) {8};
    \draw (6-\dx,0-\dy) grid (3-\dx, -1-\dy);
\end{scope}

\begin{scope}[
    every node/.append style={yslant=-0.5},
    yslant=-0.5
]
    \shade[right color=hgf-blue!10, left color=hgf-blue!10] (5,-2-\dy) rectangle +(1,5);
    \node at (5.5,2.5-\dy) {9};
    \node at (5.5,1.5-\dy) {21};
    \node at (5.5,0.5-\dy) {33};
    \node at (5.5,-0.5-\dy) {45};
    \node at (5.5,-1.5-\dy) {57};
    \draw (5,-2-\dy) grid (6,3-\dy);
     \node at (5.5, -2.5-\dy) {p$_2$}; 
\end{scope}

\begin{scope}[
    every node/.append style={yslant=0.5},
    yslant=0.5
]
    \shade[right color=hgf-blue!35,left color=hgf-blue!35] (6,-8-\dy) rectangle +(3,5);
    \node at (6.5,-3.5-\dy) {9};
    \node at (7.5,-3.5-\dy) {10};
    \node at (8.5,-3.5-\dy) {11};
    \node at (6.5,-4.5-\dy) {21};
    \node at (7.5,-4.5-\dy) {22};
    \node at (8.5,-4.5-\dy) {23};
    \node at (6.5,-5.5-\dy) {33};
    \node at (7.5,-5.5-\dy) {34};
    \node at (8.5,-5.5-\dy) {35};
    \node at (6.5,-6.5-\dy) {45};
    \node at (7.5,-6.5-\dy) {46};
    \node at (8.5,-6.5-\dy) {47};
    \node at (6.5,-7.5-\dy) {57};
    \node at (7.5,-7.5-\dy) {58};
    \node at (8.5,-7.5-\dy) {59};
    \draw (6,-8-\dy) grid (9,-3-\dy);
\end{scope}

\begin{scope}[
    every node/.append style={yslant=0.5,xslant=-1},
    yslant=0.5,
    xslant=-1
]
    \shade[bottom color=hgf-blue!22, top color=hgf-blue!22] (6-\dx,-2-\dy) rectangle +(-3,-1);
    \node at (3.5-\dx,-2.5-\dy) {9};
    \node at (4.5-\dx,-2.5-\dy) {10};
    \node at (5.5-\dx,-2.5-\dy) {11};
\draw (6-\dx,-2-\dy) grid (4-\dx, -3-\dy);
\end{scope}
\end{tikzpicture}
        \caption{\texttt{split=1}}
    \end{subfigure}
    \hspace{.5cm}
    \begin{subfigure}[b]{0.22\linewidth}
     \captionsetup{skip=-12pt}
        \begin{tikzpicture}[scale=0.47][
    every node/.style={minimum size=1cm},
    on grid

]
\begin{scope}[
    every node/.append style={yslant=-0.5},
    yslant=-0.5
]
    \shade[right color=hgf-blue!10, left color=hgf-blue!10] (0,0,-5.1945*2) rectangle +(4,5);
    \node at (0.5,4.5,-5.1945*2) {2};
    \node at (1.5,4.5,-5.1945*2) {5};
    \node at (2.5,4.5,-5.1945*2) {8};
    \node at (3.5,4.5,-5.1945*2) {11};
    \node at (0.5,3.5,-5.1945*2) {14};
    \node at (1.5,3.5,-5.1945*2) {17};
    \node at (2.5,3.5,-5.1945*2) {20};
    \node at (3.5,3.5,-5.1945*2) {23};
    \node at (0.5,2.5,-5.1945*2) {26};
    \node at (1.5,2.5,-5.1945*2) {29};
    \node at (2.5,2.5,-5.1945*2) {32};
    \node at (3.5,2.5,-5.1945*2) {35};
    \node at (0.5,1.5,-5.1945*2) {38};
    \node at (1.5,1.5,-5.1945*2) {41};
    \node at (2.5,1.5,-5.1945*2) {44};
    \node at (3.5,1.5,-5.1945*2) {47};
    \node at (0.5,0.5,-5.1945*2) {50};
    \node at (1.5,0.5 ,-5.1945*2) {53};
    \node at (2.5,0.5,-5.1945*2) {56};
    \node at (3.5,0.5,-5.1945*2) {59};
    \draw (0,0, -5.1945*2) grid (4,5,-5.1945*2);
\end{scope}

 \begin{scope}[
    every node/.append style={yslant=0.5},
    yslant=0.5
]
    \shade[right color=hgf-blue!35,left color=hgf-blue!35] (4,-8, -5.1945*2) rectangle +(1,5);
    \node at (4.5,-3.5, -5.1945*2) {11};
    \node at (4.5,-4.5, -5.1945*2) {23};
    \node at (4.5,-5.5, -5.1945*2) {35};
    \node at (4.5,-6.5, -5.1945*2) {47};
    \node at (4.5,-7.5, -5.1945*2) {59};
     \node at (4.5, -8.5, -5.1945*2) {p$_2$}; 
    \draw (4,-8, -5.1945*2) grid (5,-3, -5.1945*2);
 \end{scope}
 
 \begin{scope}[
    every node/.append style={yslant=0.5,xslant=-1},
    yslant=0.5,
    xslant=-1
]
    \shade[bottom color=hgf-blue!22, top color=hgf-blue!22] (6,1, -5.1945*2) rectangle +(-1,-4);
    \node at (5.5,0.5, -5.1945*2) {2};
    \node at (5.5,-0.5, -5.1945*2) {5};
    \node at (5.5,-1.5, -5.1945*2) {8};
    \node at (5.5,-2.5, -5.1945*2) {11};
    \draw (6,1, -5.1945*2) grid (5, -3, -5.1945*2);
\end{scope}
%
\begin{scope}[
    every node/.append style={yslant=-0.5},
    yslant=-0.5
]
    \shade[right color=hgf-blue!10, left color=hgf-blue!10] (0,0,-5.1945) rectangle +(4,5);
    \node at (0.5,4.5,-5.1945) {1};
    \node at (1.5,4.5,-5.1945) {4};
    \node at (2.5,4.5,-5.1945) {7};
    \node at (3.5,4.5,-5.1945) {10};
    \node at (0.5,3.5,-5.1945) {13};
    \node at (1.5,3.5,-5.1945) {16};
    \node at (2.5,3.5,-5.1945) {19};
    \node at (3.5,3.5,-5.1945) {22};
    \node at (0.5,2.5,-5.1945) {25};
    \node at (1.5,2.5,-5.1945) {28};
    \node at (2.5,2.5,-5.1945) {31};
    \node at (3.5,2.5,-5.1945) {34};
    \node at (0.5,1.5,-5.1945) {37};
    \node at (1.5,1.5,-5.1945) {40};
    \node at (2.5,1.5,-5.1945) {43};
    \node at (3.5,1.5,-5.1945) {46};
    \node at (0.5,0.5,-5.1945) {49};
    \node at (1.5,0.5 ,-5.1945) {52};
    \node at (2.5,0.5,-5.1945) {55};
    \node at (3.5,0.5,-5.1945) {58};
    \draw (0,0, -5.1945) grid (4,5,-5.1945);
\end{scope}
\begin{scope}[
    every node/.append style={yslant=0.5},
    yslant=0.5
]
    \shade[right color=hgf-blue!35,left color=hgf-blue!35] (4,-6, -5.1945) rectangle +(1,5);
    \node at (4.5,-1.5, -5.1945) {10};
    \node at (4.5,-2.5, -5.1945) {22};
    \node at (4.5,-3.5, -5.1945) {34};
    \node at (4.5,-4.5, -5.1945) {46};
    \node at (4.5,-5.5, -5.1945) {58};
    \node at (4.5, -6.5, -5.1945) {p$_1$}; 

    \draw (4,-6, -5.1945) grid (5,-1, -5.1945);
 \end{scope}
 \begin{scope}[
    every node/.append style={yslant=0.5,xslant=-1},
    yslant=0.5,
    xslant=-1
]
    \shade[bottom color=hgf-blue!22, top color=hgf-blue!22] (6,3, -5.1945) rectangle +(-1,-4);
    \node at (5.5,2.5, -5.1945) {1};
    \node at (5.5,1.5, -5.1945) {4};
    \node at (5.5,0.5, -5.1945) {7};
    \node at (5.5,-0.5, -5.1945) {10};
    \draw (6,3, -5.1945) grid (5, -1, -5.1945);
 \end{scope}
%
\begin{scope}[
    every node/.append style={yslant=-0.5},
    yslant=-0.5
]
    \shade[right color=hgf-blue!10, left color=hgf-blue!10] (0,0) rectangle +(4,5);
    \node at (0.5,4.5) {0};
    \node at (1.5,4.5) {3};
    \node at (2.5,4.5) {6};
    \node at (3.5,4.5) {9};
    \node at (0.5,3.5) {12};
    \node at (1.5,3.5) {15};
    \node at (2.5,3.5) {18};
    \node at (3.5,3.5) {21};
    \node at (0.5,2.5) {24};
    \node at (1.5,2.5) {27};
    \node at (2.5,2.5) {30};
    \node at (3.5,2.5) {33};
    \node at (0.5,1.5) {36};
    \node at (1.5,1.5) {39};
    \node at (2.5,1.5) {42};
    \node at (3.5,1.5) {45};
    \node at (0.5,0.5) {48};
    \node at (1.5,0.5) {51};
    \node at (2.5,0.5) {54};
    \node at (3.5,0.5) {57};
    \draw (0,0) grid (4,5);
\end{scope}
\begin{scope}[
    every node/.append style={yslant=0.5},
    yslant=0.5
]
    \shade[right color=hgf-blue!35,left color=hgf-blue!35] (4,-4) rectangle +(1,5);
    \node at (4.5,0.5) {9};
    \node at (4.5,-0.5) {21};
    \node at (4.5,-1.5) {33};
    \node at (4.5,-2.5) {45};
    \node at (4.5,-3.5) {57};
    \draw (4,-4) grid (5,1);
     \node at (4.5, -4.5) {p$_0$}; 

\end{scope}
\begin{scope}[
    every node/.append style={yslant=0.5,xslant=-1},
    yslant=0.5,
    xslant=-1
]
    \shade[bottom color=hgf-blue!22, top color=hgf-blue!22] (6,5) rectangle +(-1,-4);
    \node at (5.5,4.5) {0};
    \node at (5.5,3.5) {3};
    \node at (5.5,2.5) {6};
    \node at (5.5,1.5) {9};
    \draw (6,5) grid (5, 1);
\end{scope}
\end{tikzpicture}
        \caption{\texttt{split=2}}
    \end{subfigure}
    \caption{Distribution of a 3-D \texttt{DNDarray} across three processes: (a), \texttt{DNDarray} is not distributed, i.e., \texttt{split=None}, each process has access to the full data; (b), (c), and (d): \texttt{DNDarray} is distributed along axis 0, 1 or 2 (\texttt{split=0}, \texttt{split=1}, or \texttt{split=2}, respectively). An example for case (b) is available in \cref{lst:shape}. In each case, the data chunk labeled p$_n$ resides on process $n$, with $n = 0$, $1$, or $2$.}
    \label{fig:split_illustration}
\end{figure*}

At the core of HeAT is the Distributed N-Dimensional Array, \texttt{DNDarray} (cf. \cref{lst:shape}). The \texttt{DNDarray} object is a virtual overlay of the disjoint PyTorch tensors, which store the numerical data on each MPI process. A \texttt{DNDarray}'s data may be redundantly allocated on each node, or one-dimensionally decomposed into evenly-sized chunks with a maximum size difference of one element along the decomposition axis.
This data distribution strategy aims to balance the workload between all processes. During computation, API calls may redistribute data items. However, completed operations automatically restore the uniform data distribution. 

To steer the one-dimensional data decomposition and other parallel processing behavior, HeAT users can utilize a number of additional attributes and parameters:

\begin{itemize}
	 \item \texttt{split}: the singular axis, or dimension, along which a \texttt{DNDarray} is to be decomposed (see \cref{fig:split_illustration} and \cref{lst:shape}). If split is \texttt{None}, a redundant copy is created on each process
	 \item \texttt{device}: the computation device, i.e. CPU or GPU, on which the \texttt{DNDarray} is allocated
	 \item \texttt{comm}: the MPI communicator, i.e. the set of participating processes, for distributed computation (\cref{subsec:distributed-computation})
	 \item \texttt{shape}: the dimensionality of the global data
	 \item \texttt{lshape}: the dimensionality of the process-local data
\end{itemize}

As stated, process-level operations on \texttt{DNDarray}s are performed via PyTorch functions, thus employing their C++ core library \texttt{libtorch} to achieve high efficiency. Interoperability with external libraries such as NumPy and PyTorch is self-evident. Data contained in a NumPy \texttt{ndarray} or a PyTorch \texttt{Tensor} can be imported into a \texttt{DNDarray} via the \texttt{heat.array()} function with the optional \texttt{split} attribute. In the opposite direction, data exchange with NumPy is enabled by the \texttt{DNDarray.numpy()} method.
\begin{lstlisting}[
    caption={A \texttt{DNDarray} distributed across three processes as illustrated in \cref{fig:split_illustration}(b).},
    label={lst:shape}
]
import heat as ht
a = ht.zeros((5, 4, 3), split=0)
a.shape
[0/3] >>> (5, 4, 3)
[1/3] >>> (5, 4, 3)
[2/3] >>> (5, 4, 3)
a.lshape
[0/3] >>> (2, 4, 3)
[1/3] >>> (2, 4, 3)
[2/3] >>> (1, 4, 3)
\end{lstlisting}

A \texttt{DNDarray} can reside in a node's main memory for the CPU backend or, if available, in the VRAM of GPUs. Individual \texttt{DNDarray}s can be assigned to hardware devices via the \texttt{device} attribute or the default device can be defined as shown in \cref{lst:gpu}.
\begin{lstlisting}[
    caption={Programmatic ways of allocating \texttt{DNDarray} on different devices.},
    label={lst:gpu}
]
import heat as ht
# a single allocation
a = ht.zeros((1,), device="gpu")
a
>>> DNDarray([0.], device="gpu")

# setting a default device
ht.use_device("gpu")
b = ht.ones((1,))
b
>>> DNDarray([1.], device="gpu")
\end{lstlisting}

\subsection{Distributed Computation}
\label{subsec:distributed-computation}

Many algorithms using a distributed \texttt{DNDarray} will require communication. HeAT has a custom MPI-based communication layer composed of wrappers of point-to-point and global MPI functions. 
PyTorch's \texttt{distributed} package does not support many of the required MPI functions, such as \texttt{alltoall} or custom reduction operations. Furthermore, tensors are sent via the network as contiguous buffers in their original storage layout, so that the information on their N-dimensional structure is lost in communication. Therefore, HeAT's communication layer is based on the Python library \texttt{mpi4py}~\cite{dalcin08mpi4py}, which offers an interface to the most common MPI functions and enables the communication of contiguous Python buffer objects. 

The \texttt{DNDarray} memory representation is encoded in the one-dimensional buffer via strides (steps between elements) along the respective dimension. A main challenge in communicating an arbitrarily split \texttt{DNDarray} is the preservation of this data structure. The HeAT communication module internally handles buffer preparation as the interface between the \texttt{DNDarray} and the \texttt{mpi4py} functionality.

\begin{figure}
    \centering
    \resizebox{0.98\linewidth}{!}{ %
        \input{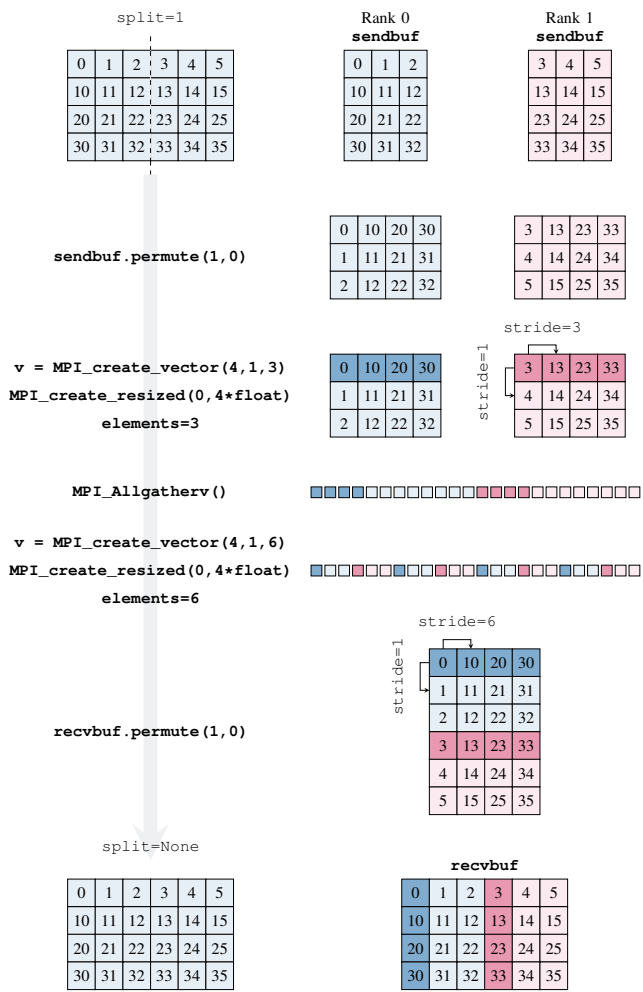}
    }
    \caption{Internal handling of a \texttt{resplit(None)} operation on a two-dimensional \texttt{DNDarray} with \texttt{split=1} in HeAT, i.e., data replication on all nodes. It depicts the on-the-fly creation of MPI datatypes for the strided access into the output and input buffers.}
    \label{fig:array-resplit}
\end{figure}

For point-to-point communications (e.g. \texttt{send, recv}), buffer preparation is trivial as the data can be sent contiguously from one process and unpacked by the receiving process. More considerable efforts must be made for communication involving collective operations. For gathering operations (e.g. \texttt{gather, allgather}), the node-local \texttt{Tensor} to be sent by each process must have the correct memory layout, which is dependent on the split axis of the \texttt{DNDarray}. For scattering operations (e.g. \texttt{scatter, alltoall}), the data chunks must be packed correctly along the split axis before distribution.

HeAT addresses the packing issues by creating custom MPI data types, which wrap the local \texttt{Tensor} buffer. First, the \texttt{DNDarray}'s dimensions are permuted such that the dimension along which data collection or distribution will take place is the first dimension. Then, custom data types are created, via the MPI function \texttt{Create\_vector}, to iteratively pack the dimensions from the last to the first. The individual data types at each dimension are defined via the \texttt{DNDarray}'s strides. The creation of such a buffer is schematically shown in \cref{fig:array-resplit}. Here, a split \texttt{DNDarray} is assembled to \texttt{split=None} via HeAT's \texttt{allgather} function.

With this internal buffer handling, HeAT offers a unified interface that provides communication without exposing the internal data representation. Based on the MPI layer, a \texttt{resplit} function is provided to change the split axis of a \texttt{DNDarray} if required. Re-splitting a \texttt{DNDarray} adheres to load balancing, i.e., the data is uniformly distributed across processes as previously stated. However, caution must be taken when using \texttt{resplit} as it is based on global MPI communication functions, thus requiring both significant communication and local memory.

In cases where CUDA-aware MPI is available, communications can be performed directly between GPUs. Otherwise, data must be copied from the GPU to the CPU, sent to the target CPU, then copied to the target GPU. This increases the communication overhead, and therefore the run time, of many functions. 

\section{Performance Results}
\label{sec:performance-result}

Performance of HeAT was evaluated by benchmarking algorithms that are commonly used in data science, and comparing the results to implementations in other frameworks. NumPy and PyTorch were chosen for single-node baseline evaluation. NumPy because it is the most frequently used library for Python-based array computation, and PyTorch because it is the underlying node-local eager execution engine for HeAT. For multi-node experiments, Dask was selected. 

Algorithms were chosen based on their algorithmic complexity and availability as ready-to-use implementations in the investigated frameworks. Two types of low-level algorithms were benchmarked: the computation of statistical moments (\cref{subsubsec:performance-mean}) and the computation of pairwise Euclidean distances, i.e. the function \texttt{cdist} (\cref{subsubsec:performance-cdist}). These operations are available in Dask, PyTorch, and NumPy as designated functions under the same name and mathematical definition. 
As HeAT aims to provide both low-level and high-level functionality, two commonly used ML algorithms were also selected: \textit{k}-means clustering (\cref{subsubsec:performance-kmeans}) and least absolute shrinkage and selection operator (LASSO) regression (\cref{subsubsec:performance-lasso}). These two algorithms are provided by Dask and scikit-learn as end-user implementations. For comparison with PyTorch, algorithms were specifically implemented with as much provided PyTorch functionality as possible. All benchmark scripts are available on HeAT's GitHub repository.

\subsection{Execution Environment}
\label{subsec:execution-environment}

\begin{table}[h]
\caption{Software packages used for performance benchmarks.}
\begin{center}
\begin{tabular}{|l l|l l|}
\hline
\multicolumn{2}{|l|}{\textbf{General}}&\multicolumn{2}{l|}{\textbf{Python}} \\
\textbf{\textit{Package}} & \textbf{\textit{Version}}& \textbf{\textit{Package}}& \textbf{\textit{Version}} \\
\hline
CUDA & 10.2& dask  &  2.12.0 \\
GCC & 8.3.0 &dask-ml & 1.2.0 \\
HDF5 & 1.10.5 & dask-mpi & 2.0.0 \\
Intel Cluster Studio XE & 2019.03 &heat  &  0.4.0 \\
ParaStationMPI &  5.2.2-1 & mpi4py & 3.0.3 \\  
Python & 3.6.8 & numpy$^{\mathrm{a}}$ &  1.15.2 \\
  & & sklearn & 0.22.2 \\
  & &torch &1.5.0 \\
\hline
\multicolumn{4}{l}{$^{\mathrm{a}}$using \texttt{Intel MKL 2019.1}.}
\end{tabular}
\label{tab:software-env}
\end{center}
\end{table}

The software environment for benchmarking is summarized in \cref{tab:software-env}. The experiments were run on a machine learning HPC system comprised of 15 compute nodes with commodity components at the Jülich Supercomputing Centre (JSC). Each node is equipped with two 12-core Intel Xeon Gold 6126 CPUs, \SI{206}{\giga\byte} of DDR3 main memory, and four NVIDIA Tesla V100 SXM2 GPUs with \SI{32}{\giga\byte} VRAM per card. The GPUs communicate node-internally via an NVLink interconnect. The system is optimized for GPUDirect communication across node boundaries, supported by 2x Mellanox \SI{100}{\giga\bit} EDR InfiniBand links. Though HeAT supports CUDA-aware MPI, it was not used, as to make experiments comparable to Dask, which cannot make use of this. MPI capable commodity clusters should show similar differences between HeAT and Dask unless the cluster is specifically tuned for non-standard use cases. 

\subsection{Datasets}
\label{subsubsec:datases}

Three datasets were chosen to demonstrate the effectiveness of HeAT for different data characteristics and to mimic common use-cases:
\begin{enumerate}
    \item The Cityscapes dataset~\cite{cordts2016cityscapes} contains 5\:000 high-resolution images with fine-grained annotations. Each image is 2\:048$\times$1\:024 pixels with three 256-bit RGB color channels per pixel, which have been flattened and combined into a short-fat matrix with 5\:000$\times$6\:291\:456 entries, i.e. \SI{117.19}{\giga\byte}. 
    \item The SUSY dataset~\cite{baldi2014searching} contains 5\,000\,000 samples from Monte Carlo simulations of high-energy particle collisions. Each sample has 18 features, consisting of kinematic properties measured by particle detectors and high-level functional derivations of those measurements~\cite{uciMachineLearningRepository}. The total data size is \SI{343.33}{\mega\byte}
    \item The EURopean Air pollution Dispersion-Inverse Model data (EURAD-IM)~\cite{elbern20004dvar} contains parameters from an Eulerian meso-scale chemistry transport model as part of the Copernicus Atmosphere Monitoring Service (CAMS)\footnote{https://atmosphere.copernicus.eu/}. For our experiments, $10^7$ data points and 100 parameters, i.e. \SI{7.45}{\giga\byte}, of the model have been chosen and stored in a tall-skinny matrix.
\end{enumerate}
The Cityscapes and SUSY datasets are publicly available, the EURADS dataset is available upon request. All datasets were converted from their original sources into solitary data matrices and stored as floating point values in HDF5 files~\cite{hdf52020hdf5}. 

While both Dask and HeAT utilize parallel I/O for \texttt{h5py}~\cite{collette2014h5py}, they handle data decomposition differently. HeAT automatically decomposes data upon \texttt{DNDarray} creation when given a split axis by the user (cf. \cref{subsec:dndarrays}). Dask offers an automatic data decomposition scheme, the recommended setup as per the Dask documentation \cite{daskdocumentation}, and manually specifying the size of the memory-distributed data chunks. In the following, Dask's performance with automatic chunking is indicated with Dask-auto, whereas Dask-tuned indicates the manual chunking mirroring HeAT's data decomposition scheme. All measurements with HeAT are performed with \texttt{split=0}, unless otherwise stated.

\subsection{Experiments}
\label{subsec:applications}
 
All of the following experiments are composed of weak scaling and strong scaling experiments. Measurements are the average of 9 runs, preceded by a warm up run. The error bars indicate the empirical standard deviation. In several cases the errors are too small to be visible compared to the data point itself. A fractional number of nodes for weak scaling GPU runs refers to the usage of the equivalent fraction of a node's resources, i.e. one (0.25) or two (0.5) out of the nodes four GPUs.

For Dask, the actual program code is provided in a separate script that connects the scheduler to the workers via a \texttt{dask.distributed.Client} instance. The discovery of the scheduler is done manually by passing an IP address, or via information on a shared filesystem. Networking between the processes builds on network sockets and utilizes Infiniband using TCP over IB. Each worker maintains its execution state by writing into journaling directories. 

Weak scaling refers to the process of increasing the amount of computation resources while maintaining the workload on each MPI process. Ideal weak scaling behaviour is a constant runtime for each measurement as the number of processing units is increased. This indicates solid scalability for larger datasets. Results for weak scaling experiments are presented as the average maximum runtime across the processing units.

Strong scaling refers to the process of increasing the amount of computation resources while the total workload on the system remains constant. Ideally, the runtime in strong scaling measurements is inversely proportional to the number of processing units. We present our strong scaling results in units of speedup as compared to a single-node NumPy-based implementation. 

\subsubsection{Statistical Moments}
\label{subsubsec:performance-mean}

\begin{figure*}[tb]
    \centering
    \begin{subfigure}[c]{\textwidth}
    	\input{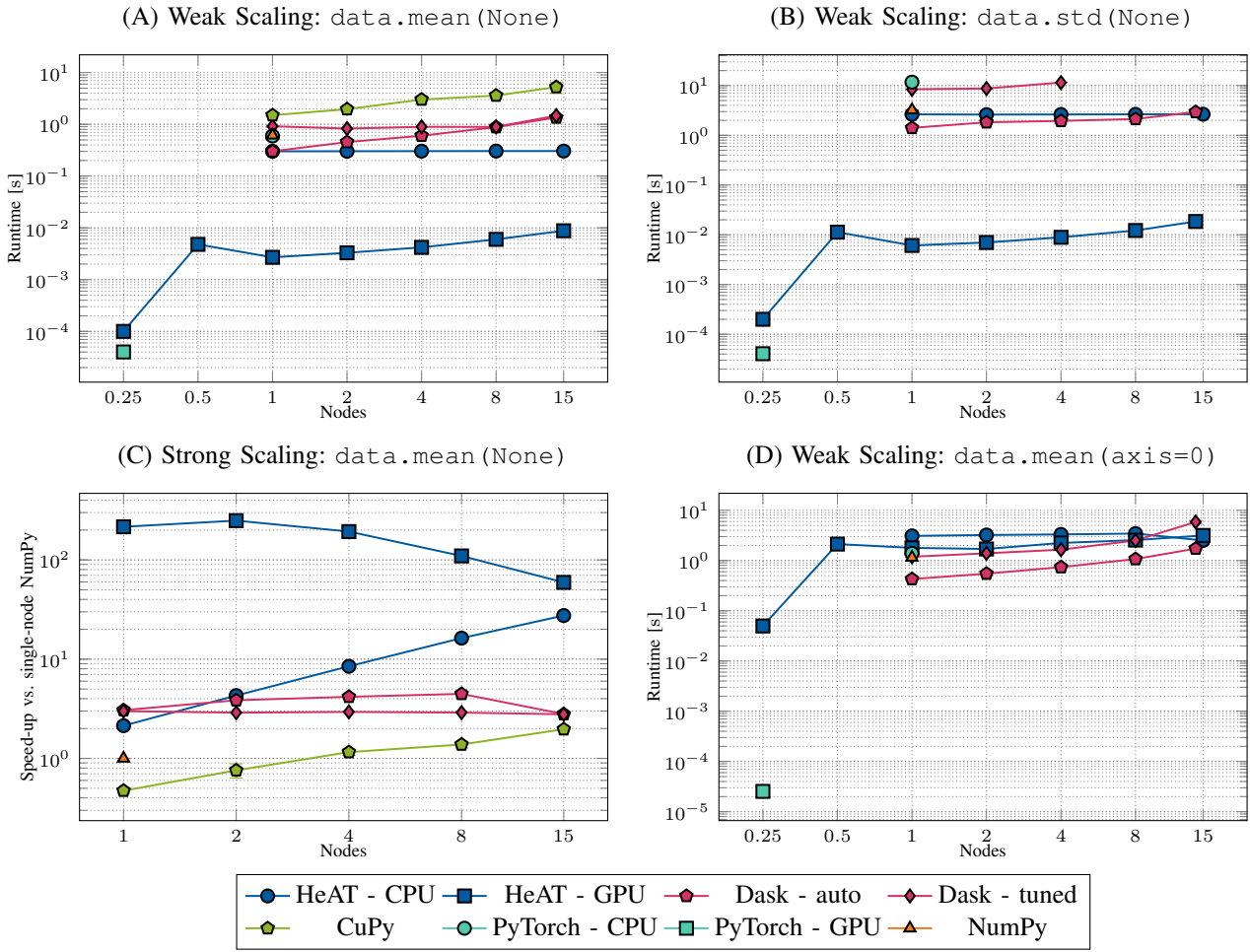}
    \end{subfigure}      
    \caption{Mean and standard deviation measurements of the Cityscapes dataset. (A) weak scaling of \texttt{mean(data, axis=None)}, (B) weak scaling of \texttt{std(data,axis=None)}, (C) strong scaling of \texttt{mean(data, axis=None)}, (D) weak scaling of \texttt{mean(data, axis=0)}. Cf. \cref{subsubsec:performance-mean}. 0.25 and 0.5 nodes refer to usage of only one and two out of the node's four GPUs.}
    \label{fig:scaling-moments}
\end{figure*}

Calculation of mean and standard deviation are arguably the most frequently used calculations in all of computing. In a distributed context, it is inefficient to compute statistical moments with multiple passes over the dataset. Therefore, HeAT calculates statistical moments using the numerically stable single-pass algorithms presented in \cite{moments}. These experiments utilise the Cityscapes dataset.

The experiments shown in this section show some of the most common applications of statistical moments: the mean of the entire dataset, the mean along the largest dimension, and the standard deviation of the entire dataset. In its native form, Dask is designed for distributed computation on CPUs. Multi-node GPU usage can be enabled by coupling of Dask to CuPy. In order to provide comparison of HeAT's multi-GPU performance, benchmarks for Dask were originally intended to be performed with and without CuPy. However, during benchmarks we found that the native binding of CuPy to Dask only enables the usage of a single GPU per node, while HeAT enables multi-GPU per node usage. While possible for Dask to utilize multiple GPUs per node, substantial modifications to each algorithm requiring communication must be done manually by the user. Therefore multi-GPU benchmarks on Dask with CuPy were only performed for mean calculations of the entire dataset. Exemplary code for the algorithmic definition can be seen in the \cref{lst:DaskCupy} (Appendix).

\cref{fig:scaling-moments} shows strong and weak scaling of the mean operation, the weak scaling of standard deviation, and the weak scaling of the mean along the largest axis. For weak scaling runs, each node had 300 rows of the matrix. For strong scaling runs, 1\:200 rows were used globally. Dask-tuned failed with memory errors above 4 nodes for the standard deviation calculations. Single-GPU measurements were much faster than multi-GPU measurements as no communication is required. Furthermore, PyTorch is faster than HeAT on a single GPU as the HeAT functions are wrappers of PyTorch functions. Favorable strong scaling is clearly shown in the HeAT CPU measurements (\cref{fig:scaling-moments} (C)). HeAT outperforms NumPy and is nearly equal to Dask-auto, for a single node. However, beyond one node, HeAT clearly improves upon these results, while all Dask measurements do not show significant improvement with the addition of more computing resources. CuPy shows positive scaling behavior, however it is the slowest of the three parallel frameworks. As the number of nodes increases, the time required for communication also increases. Once this time eclipses the time required for computation the performance degrades. This effect can be seen for the HeAT GPU measurements beyond four nodes. The effect is common in distributed computing as one must balance the number of MPI processes and the size of the dataset to avoid excessive communication calls. 

HeAT calculations on CPU for these experiments show nearly ideal weak scaling. However, the measurements of the mean along the largest axis show Dask outperforming HeAT until the number of nodes is increased beyond 8. This is due to the large difference between the complexity of the calculation and the amount of communication required by HeAT. As the complexity of the calculation increases, the efficiency of HeAT remains constant and at 15 nodes shows slight improvement. Whereas the efficiency of Dask degrades as the node count increases.

All of these measurements were also conducted with the \texttt{split=1} data distributed scheme for HeAT. The difference between the runtimes for \texttt{split=0} and \texttt{split=1} was on average \SI{-0.0226 \pm 0.0809}{\second}. 

\subsubsection{Pairwise Euclidean Distances}
\label{subsubsec:performance-cdist}

Pairwise distance calculations are a vital part of data analysis  used in many ML algorithms, such as clustering and neighborhood methods~\cite{debus2020high}. However, due to the quadratic growth in computational complexity, these computations scale notoriously poorly. Moreover, their excessive memory consumption poses a major challenge, which often limits the number of samples which can be processed. As a consequence, many applications employ a form of dimensionality reduction, yielding only approximate solutions.
HeAT implements a custom distance computation function via ring communication which works regardless of the employed distance metric. 

\begin{figure}[tb]
    \centering
    \hspace{-0.45cm}
    \begin{subfigure}[c]{0.44\textwidth}
    	\begin{tikzpicture}
    \begin{axis}[
        BaseAxis,
        xlabel={Nodes},
        xmode=log,
        log basis x=2,
        xticklabel=\pgfmathparse{2^\tick}\pgfmathprintnumber{\pgfmathresult},
        x label style={font=\scriptsize, yshift=0.3em},
        xtick={0.25, 0.5, 1, 2, 4, 8, 15},
        ymode=log,
        ylabel={Runtime, [\si{\second}]},
        ymin=1e-4, 
        ymax=1e3,
        max space between ticks=20
    ]
    \addplot[color=hgf-blue, mark=*, BasePlot]
    table [x=x, y=y, y error=y-err]{%
        x   y           y-err
        1   4.07184     0.06414
        2   4.70951     0.31232
        4   6.08104     0.02579
        8   10.19623    0.08235
        15  20.34080    0.11117
    };
    \addplot[color=hgf-blue, mark=square*, BasePlot]
    table [x=x, y=y, y error=y-err]{%
        x       y           y-err
        0.25    0.00040     0.00001
        0.5     0.01409     0.00028
        1       0.01727     0.00007
        2       0.02271     0.00009
        4       0.03408     0.00007
        8       0.05856     0.00200
        15      0.09617     0.00104
    };
    \addplot[color=hgf-red, mark=*, BasePlot]
    table [x=x, y=y, y error=y-err]{%
        x   y           y-err
        1   28.353      0.602
        2   55.401      0.499
        4   111.262     0.364
        8   369.140     44.318
   };
    \addplot[color=hgf-red, mark=diamond*, BasePlot]
    table [x=x, y=y, y error=y-err]{%
        x   y           y-err
        1   27.434      0.195
        2   60.024      1.145
        4   116.646     1.63
        8   210.207     24.93
    };
    \addplot[color=hgf-turqoise, mark=*, BasePlot]
    table [x=x, y=y, y error=y-err]{%
        x   y                   y-err
        1   2.571144327         0.520167367
    };
    \addplot[color=hgf-turqoise, mark=square*, BasePlot]
    table [x=x, y=y, y error=y-err]{%
        x       y               y-err
        0.25    0.000633356     0.001131427
    };
    \addplot[color=hgf-orange, mark=triangle*, BasePlot]
    table [x=x, y=y, y error=y-err]{%
        x   y               y-err
        1   4.925835746     0.338843744
    };
    \end{axis}
\end{tikzpicture}
    \end{subfigure}      
    \begin{subfigure}[c]{0.44\textwidth}
        \begin{tikzpicture}
    \begin{axis}[
        BaseAxis,
        xlabel={Nodes},
        xmode=log,
        log basis x=2,
        xticklabel=\pgfmathparse{2^\tick}\pgfmathprintnumber{\pgfmathresult},
        x label style={font=\scriptsize, yshift=0.3em},
        xtick={0.5, 1, 2, 4, 8, 15},
        ymode=log,
        ylabel={Speedup vs. single-node NumPy},
        legend columns=4,
        legend style={at={(0.5,-0.175)},anchor=north},
        ymin=1e-3, 
        ymax=1e3,
        max space between ticks=20
    ]
    \addplot[color=hgf-blue, mark=*, BasePlot]
    table [x=x, y=y, y error=y-err]{%
        x   y                   y-err
        1   1.34785         0.06869
        2   2.28709         0.11742
        4   2.74999         0.08367
        8   1.58895         0.07197
        15  0.71295         0.07051
    };
    \addlegendentry{HeAT, CPU};
    \addplot[color=hgf-blue, mark=square*, BasePlot]
    table [x=x, y=y, y error=y-err]{%
        x       y               y-err
        1       343.46509        0.06875
        2       469.62080        0.07703
        4       448.81436        0.07439
        8       273.60860        0.15315
        15      185.94702        0.07643
    };
    \addlegendentry{HeAT, GPU};
    \addplot[color=hgf-red, mark=pentagon*, BasePlot]
    table [x=x, y=y, y error=y-err]{%
        x   y                   y-err
        1   0.077234801         0.072902585
        2   0.169669792         0.07146952
        4   0.171354294         0.080798222
        8   0.170478677         0.080983766
        14  0.173194499         0.079802111
    };
    \addlegendentry{Dask, auto};
    \addplot[color=hgf-red, mark=diamond*, BasePlot]
    table [x=x, y=y, y error=y-err]{%
        x   y                   y-err
        1   0.172929397         0.077805813
        2   0.167036034         0.090690013
        4   0.163453484         0.07582409
        8   0.313360588         0.088273892
        14  0.477741511         0.100242146
    };
    \addlegendentry{Dask, tuned};
    \addlegendimage{color=hgf-turqoise, mark=*, BasePlot}
    \addlegendentry{PyTorch, CPU};
    \addlegendimage{color=hgf-turqoise, mark=square*, BasePlot}
    \addlegendentry{PyTorch, GPU};
    \addplot[color=hgf-orange, mark=triangle*, BasePlot]
    table [x=x, y=y, y error=y-err]{%
        x       y                   y-err
        1       1.000000000         0.0
    };
    \addlegendentry{NumPy};
    \end{axis}
\end{tikzpicture}
    \end{subfigure}   
    \caption{Pairwise euclidean distances: weak (\emph{upper}) and strong scaling measurements (\emph{lower}), cf. \cref{subsubsec:performance-cdist}. 0.25 and 0.5 nodes refer to usage of only one and two out of the node's four GPUs.}
    \label{fig:scaling-cdist}
\end{figure}
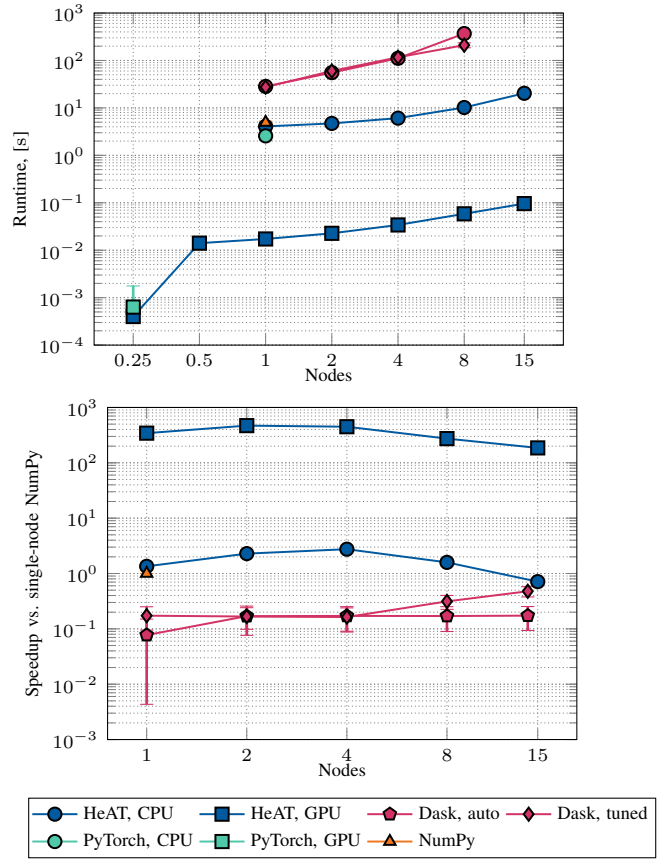

For scaling experiments, the $L_2$ norm (Euclidean distance, commonly referred to as \texttt{cdist}) was utilized. Benchmarks were conducted on the SUSY dataset. For weak scaling, the number of samples was increased by the square root of the number of nodes, because the computational load grows quadratically. The first 12\:910, 18\:258, 25\:820, 36\:515, 51\:640, 73\:030, and 100\:000 samples were used for $N=$ 0.25, 0.5, 1, 2, 4, 8 and 15 nodes, respectively. For strong scaling runs, the first 40\:000 samples were used.
Results are displayed in \cref{fig:scaling-cdist}, where HeAT's implementation of \texttt{cdist} shows significantly lower computation times compared to the other frameworks. It also provides small speedup for CPU and large speedup for GPU over the NumPy implementation. On one GPU (0.25 nodes), HeAT outperforms PyTorch because it employs quadratic expansion via matrix multiplication for the calculation of the squared differences rather than relying on PyTorch's \texttt{cdist} function:
\begin{equation}
    \|X - Y\|_2^2 = \|X\|_2^2 + \|Y\|_2^2 - 2\;X\cdot Y
\end{equation}
On CPU, PyTorch's intrinsic \texttt{cdist} function is faster; however, experiments for \textit{k}-means, cf. \cref{subsubsec:performance-kmeans}, will show that this speedup depends on the data matrix shape (tall-skinny vs short-fat).

Overall scaling behaviour of the function is  not optimal, as the communication overhead grows proportionally to the number of processes. Nevertheless, the HeAT implementation is able to solve the problem of memory usage at large sample sizes, whereas Dask's computation at 14 Nodes with 100\:000 samples aborted due to memory overflow.

\subsubsection{\textit{k}-means}
\label{subsubsec:performance-kmeans}

\textit{k}-means \cite{macqueen1967kmeans} is a vector quantization method originating from the field of signal processing. It is commonly used as an unsupervised clustering algorithm that is able to assign all observations, $x$, within a dataset into $k$ disjoint partitions, each forming a cluster, ($C_i$). Formally, the method minimizes the inter-cluster variance:

\begin{equation}
\argmin_{C}\sum_{i=1}^{k}\sum_{x\in C_i}\|x - c_i\|_2^2
\end{equation}

for each cluster centroid $c_i$. The \textit{k}-means clustering problem is generally NP-hard, but can be efficiently approximated using an iterative optimization method, such as detecting a local minimum, i.e. Lloyd's algorithm~\cite{lloyd1982least}. HeAT's \textit{k}-means implementation is dominated by element-wise vector operations in the distance matrix computation between the data points and the centroids, and reduction operations for finding the best matching centroids. Benchmarks in this experiment were conducted on the Cityscapes dataset. The dataset sizes used were analogous to those used in the moments experiments. For each benchmark, we have performed 30 iterations of Lloyd's algorithm at eight assumed centroids.

\begin{figure}[tb]
    \centering
    \begin{subfigure}[c]{0.44\textwidth}
    	\begin{tikzpicture}
    \begin{axis}[
        BaseAxis,
        xlabel={Nodes},
        xmode=log,
        log basis x=2,
        xticklabel=\pgfmathparse{2^\tick}\pgfmathprintnumber{\pgfmathresult},
        x label style={font=\scriptsize, yshift=0.3em},
        xtick={0.25, 0.5, 1, 2, 4, 8, 15},
        ymode=log,
        ylabel={Runtime, [\si{\second}]},
    ]
    \addplot[color=hgf-blue, mark=*, BasePlot]
    table [x=x, y=y, y error=y-err]{%
        x   y                   y-err
        1   89.9710261083932    1.2458840356568
        2   92.0141701855593    1.25776838102127
        4   97.0548371001044    2.72889144165486
        8   99.0493648626532    0.929681498502406
        15  106.998394381565    1.01261258371691
    };
    \addplot[color=hgf-blue, mark=square*, BasePlot]
    table [x=x, y=y, y error=y-err]{%
        x       y                   y-err
        0.25    2.30535929008491    0.000247639532841
        0.5     7.1423864149385     0.002110125928383
        1       8.62856516987085    0.003739979359372
        2       9.45222009110472    0.004606403428391
        4       9.97552788269726    0.011980117572718
        8       10.3344660657975    0.013102308490763
        15      13.0280166059318    0.006371339036347
    };
    \addplot[color=hgf-red, mark=*, BasePlot]
    table [x=x, y=y, y error=y-err]{%
        x   y                   y-err
        1   436.224853881634    8.44477019025915
        2   471.549366125837    0
    };
    \addplot[color=hgf-red, mark=diamond*, BasePlot]
    table [x=x, y=y, y error=y-err]{%
        x   y                   y-err
        1   190.439132806949    0.802554498367578
        2   230.811883910456    1.30737587250231
        4   461.090210907161    0
    };
    \addplot[color=hgf-turqoise, mark=*, BasePlot]
    table [x=x, y=y, y error=y-err]{%
        x   y                   y-err
        1   188.633089635521    0.303897506642439
    };
    \addplot[color=hgf-turqoise, mark=square*, BasePlot]
    table [x=x, y=y, y error=y-err]{%
        x       y                   y-err
        0.25    11.4632823267538    0.000770101634984
    };
    \addplot[color=hgf-orange, mark=triangle*, BasePlot]
    table [x=x, y=y, y error=y-err]{%
        x   y                   y-err
        1   2584.38244208435    5.5251633023935
    };
    \end{axis}
\end{tikzpicture}
    \end{subfigure}      
    \begin{subfigure}[c]{0.44\textwidth}
	 \begin{tikzpicture}
    \begin{axis}[
        BaseAxis,
        xlabel={Nodes},
        xmode=log,
        log basis x=2,
        xticklabel=\pgfmathparse{2^\tick}\pgfmathprintnumber{\pgfmathresult},
        x label style={font=\scriptsize, yshift=0.3em},
        xtick={0.5, 1, 2, 4, 8, 15},
        ymode=log,
        ylabel={Speedup vs. single-node NumPy},
        legend columns=4,
        legend style={at={(0.5,-0.175)},anchor=north}
    ]
    \addplot[color=hgf-blue, mark=*, BasePlot]
    table [x=x, y=y, y error=y-err]{%
        x   y                   y-err
        1   29.6883116128879    0.899780295093525
        2   55.0748954819056    0.883627230608143
        4   105.113384712341    1.56594395906776
        8   169.060727451777    2.28105100255988
        15  186.738021421356    1.54604142141775
    };
    \addlegendentry{HeAT, CPU};
    \addplot[color=hgf-blue, mark=square*, BasePlot]
    table [x=x, y=y, y error=y-err]{%
        x   y                   y-err
        1   606.681051112       0
        2   567.965765244       0
        4   584.965799023       0
        8   589.070453913       0
        15  553.400428436       0
    };
    \addlegendentry{HeAT, GPU};
    \addplot[color=hgf-red, mark=*, BasePlot]
    table [x=x, y=y, y error=y-err]{%
        x   y                   y-err
        2   0.561959520688355   0
        4   0.704349638466698   0.007143472244869
        8   0.477586530521428   0
        14  0.573612866658415   0
    };
    \addlegendentry{Dask, auto};
    \addplot[color=hgf-red, mark=diamond*, BasePlot]
    table [x=x, y=y, y error=y-err]{%
        x   y                   y-err
        1   4.87689614752499    0.019835441718687
        2   8.26460498288493    0.024003362845877
        4   8.36043117106889    0.026193123398747
        8   5.6735506097246     0.035389018828569
    };
    \addlegendentry{Dask, tuned};
    \addlegendimage{color=hgf-turqoise, mark=*, BasePlot}
    \addlegendentry{PyTorch, CPU};
    \addlegendimage{color=hgf-turqoise, mark=square*, BasePlot};
    \addlegendentry{PyTorch, GPU};

    \addplot[color=hgf-orange, mark=triangle*, BasePlot]
    table [x=x, y=y, y error=y-err]{%
        x   y   y-err
        1   1   0.001692912026329
    };
    \addlegendentry{NumPy};
    \end{axis}
\end{tikzpicture}
   \end{subfigure}   
    \caption{\textit{k}-means clustering: weak (\emph{upper}) and strong scaling measurements (\emph{lower}), cf. \cref{subsubsec:performance-kmeans}. 0.25 and 0.5 nodes refer to usage of only one and two out of the node's four GPUs.}
    \label{fig:scaling-kmeans}
\end{figure}
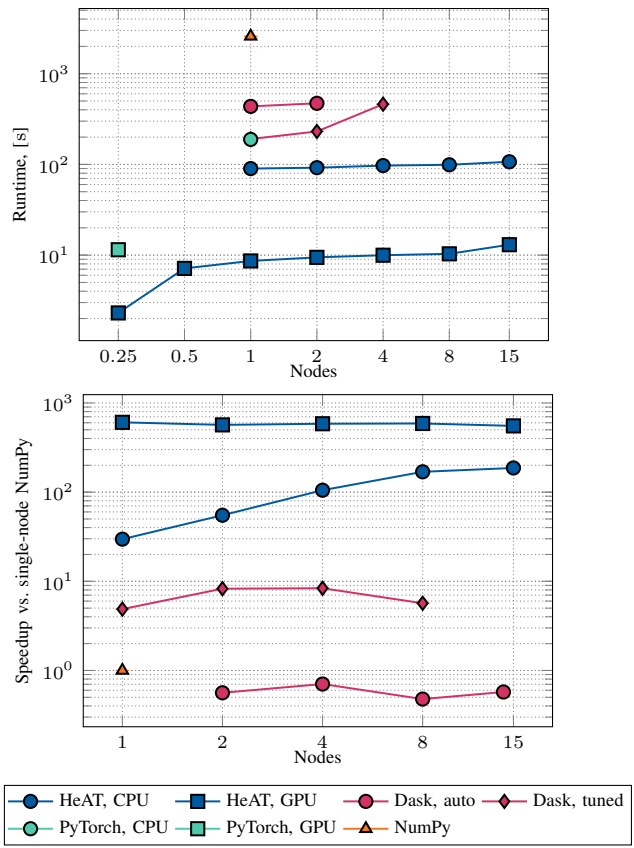

Weak scaling measurements in \cref{fig:scaling-kmeans} (upper panel) show that HeAT outperforms Dask by at least an order of magnitude. Furthermore, HeAT demonstrates solid scalability on both CPU and GPU. For Dask we were unable to complete the measurement procedure for all node configurations. While it was sporadically possible to complete the benchmark with a four-node configuration, it would terminate with an out-of-memory exception before the completion of the measurement sequence. For 8 and 15 nodes, we were unable to obtain any measurements due to excessive memory consumption.

For HeAT, a single GPU shows better overall performance compared to multiple GPUs. The difference in runtime between PyTorch and HeAT on a single node can be explained by the differences in distance matrix computation (c.f. \cref{subsubsec:performance-cdist}).
Strong scaling measurements are shown in \cref{fig:scaling-kmeans} (lower panel). For these measurements, 600 rows of the dataset were used for all runs. Here, we obtain similar conclusions as in the weak scaling measurements. HeAT outperforms Dask by a significant margin and shows more favorable scaling behaviour. Again, Dask experienced out-of-memory issues. While HeAT's CPU computations scale approximately linearly, the GPU backend shows strong linearity.
\newline

\subsubsection{LASSO}
\label{subsubsec:performance-lasso}

LASSO is a regression method of simultaneously applying regularization and parameter selection. Its basic form is an extension of the ordinary linear regression (OLS) method by introducing an L1-norm penalty of the parameters scaled by the regularization parameter. The corresponding objective function reads

\begin{equation}
\label{equ:lossfkt}
E(w) = \|y - X w \|_2^2 + \lambda \|w_{-} \|_1
\end{equation}

where $y$ denotes the $n$ samples of the output variables, $X \in \mathbb{R}^{n \times m}$ denotes the \emph{system matrix} in which $m-1$ columns represent the different features in which each column represents the constant bias term and each of the $n$ rows represent one data sample, $w \in \mathbb{R}^{m}$ denotes the regression coefficients,  $w_{-} \in \mathbb{R}^{m-1}$ the regression coefficients of the features, and $\lambda$ the regularization parameter.
In addition to the L2-norm regularization approach (i.e., Ridge-regression), LASSO favors not only smaller model parameters but, depending on the regularization parameters, can force selected model parameters to be zero. It is a popular method to determine the importance of input variables with respect to one or more dependent output variables. 

In this experiment, a LASSO algorithm is used to determine the most important model parameters of the EURAD-IM model on the errors of ozone forecasts of the model at measurement sites\footnote{obtained from the centralised AirBase database 
maintained by the European Environment Agency (EEA). The AirBase database collects
near real time data from the European countries bound under Decision 97/101/EC engaging
exchange of information  on ambient air quality.}. In order to minimize the objective function, a coordinate descent algorithm with a proximal gradient soft threshold applied to each coordinate was implemented in HeAT, Dask, NumPy, and PyTorch. For the weak scaling measurements, the LASSO algorithm is run for 20 iterations on a data sample size of 714\:280 samples per node.

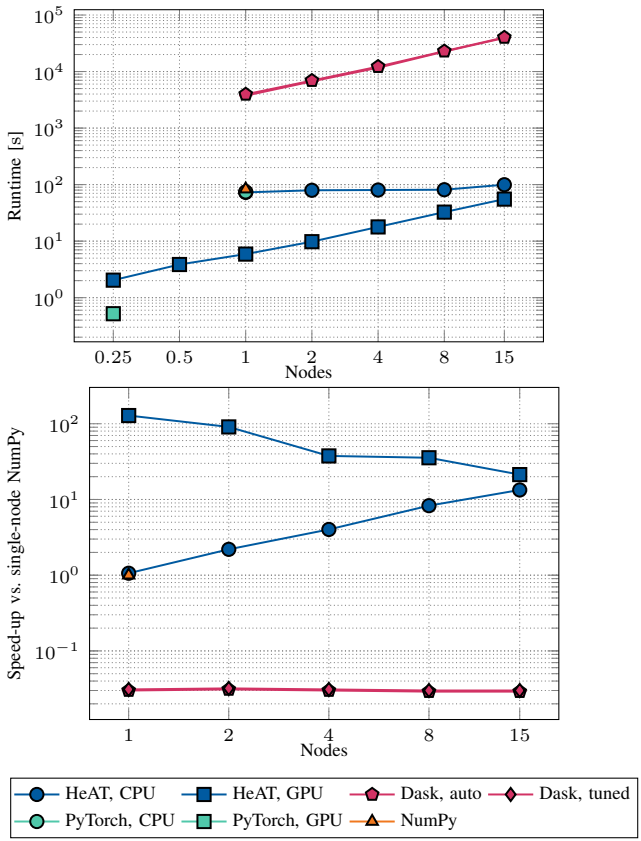
\begin{figure}[tb]
    \centering
    \begin{subfigure}[c]{0.44\textwidth}
    	\begin{tikzpicture}
    \begin{axis}[
        BaseAxis,
        xlabel={Nodes},
        xmode=log,
        log basis x=2,
        xticklabel=\pgfmathparse{2^\tick}\pgfmathprintnumber{\pgfmathresult},
        x label style={font=\scriptsize, yshift=0.3em},
        xtick={0.25, 0.5, 1, 2, 4, 8, 15},
        ymode=log,
        ylabel={Runtime [\si{\second}]},
    ]
    \addplot[color=hgf-blue, mark=*, BasePlot]
    table [x=x, y=y, y error=y-err]{%
        x   y    y-err
        1   72.80   0.374      
        2   79.40   0.116
        4   80.20   0.296   
        8   81.20   0.326     
        15  99.60   0.488
    };
    \addplot[color=hgf-blue, mark=square*, BasePlot]
    table [x=x, y=y, y error=y-err]{%
        x   y                   y-err
        0.25 2.04            0.022
        0.5  3.86            0.028
        1    5.90            0.072
        2    9.78            0.084
        4    17.84           0.22
        8    32.60           0.30
        15   55.40           3.41
    };
    \addplot[color=hgf-red, mark=diamondpentagon*, BasePlot]
    table [x=x, y=y, y error=y-err]{%
        x   y                   y-err
        1   3742     6.2
        2   6688     8.6
        4   11674    4.2
        8   22544    17.4
        15  39308    22.0
    };
    \addplot[color=hgf-red, mark=pentagon*, BasePlot]
    table [x=x, y=y, y error=y-err]{%
        x   y                   y-err
        1   3956     1.6
        2   6922     22    
        4   12222    20
        8   22930    56
        15  40140    84
    };
    \addplot[color=hgf-turqoise, mark=*, BasePlot]
    table [x=x, y=y, y error=y-err]{%
        x   y       y-err
        1   74.20   0.17
    };
    \addplot[color=hgf-turqoise, mark=square*, BasePlot]
    table [x=x, y=y, y error=y-err]{%
        x       y           y-err
        0.25    0.52      0.0086 
    };
    \addplot[color=hgf-orange, mark=triangle*, BasePlot]
    table [x=x, y=y, y error=y-err]{%
        x   y       y-err
        1   83.20   0.75
    };
    \end{axis}
\end{tikzpicture}
    \end{subfigure}      
    \begin{subfigure}[c]{0.44\textwidth}
	 \begin{tikzpicture}
    \begin{axis}[
        BaseAxis,
        xlabel={Nodes},
        xmode=log,
        log basis x=2,
        xticklabel=\pgfmathparse{2^\tick}\pgfmathprintnumber{\pgfmathresult},
        x label style={font=\scriptsize, yshift=0.3em},
        xtick={0.5, 1, 2, 4, 8, 15},
        ymode=log,
        ylabel={Speed-up vs. single-node NumPy},
        legend columns=4,
        legend style={at={(0.5,-0.175)},anchor=north}
    ]
    \addplot[color=hgf-blue, mark=*, BasePlot]
    table [x=x, y=y, y error=y-err]{%
        x   y        y-err
        1  1.0611    0.0211
        2  2.1952    0.0513
        4  4.0183    0.0747
        8  8.2791    0.2008
        15 13.355    0.3662
    };
    \addlegendentry{HeAT, CPU};
    \addplot[color=hgf-blue, mark=square*, BasePlot]
    table [x=x, y=y, y error=y-err]{%
        x   y        y-err
        1   128.32  1.77
        2   90.815  3.17
        4   37.59   1.42
        8   35.56   1.50  
        15  21.31   0.75
    };
    \addlegendentry{HeAT, GPU};
    \addplot[color=hgf-red, mark=pentagon*, BasePlot]
    table [x=x, y=y, y error=y-err]{%
        x   y   y-err
        1  0.03    0.0002
        2  0.031   0.0002
        4  0.030   0.0002
        8  0.029   0.0002
        15 0.029   0.0002
    };
    \addlegendentry{Dask, auto};
    \addplot[color=hgf-red, mark=diamond*, BasePlot]
    table [x=x, y=y, y error=y-err]{%
        x   y   y-err
        1  0.031   0.0002
        2  0.032   0.0003
        4  0.031   0.0002
        8  0.030   0.0002
        15 0.030   0.0002
    };
    \addlegendentry{Dask, tuned};
    \addlegendimage{color=hgf-turqoise, mark=*, BasePlot}
    \addlegendentry{PyTorch, CPU};
    \addlegendimage{color=hgf-turqoise, mark=square*, BasePlot}
    \addlegendentry{PyTorch, GPU};
    \addplot[color=hgf-orange, mark=triangle*, BasePlot]
    table [x=x, y=y, y error=y-err]{%
        x   y   y-err
        1   1   0.0146
    };
    \addlegendentry{NumPy};
    
     \end{axis}
\end{tikzpicture}
   \end{subfigure}   
    \caption{LASSO regression: weak (\emph{upper}) and strong scaling measurements (\emph{lower}), cf. \cref{subsubsec:performance-lasso}. 0.25 and 0.5 nodes refer to usage of only one and two out of the node's four GPUs.}
    \label{fig:scaling-lasso}
\end{figure}

The HeAT CPU measurements show good weak scaling behaviour (\cref{fig:scaling-lasso}, upper panel) with the lowest runtime compared to the Dask and HeAT GPU versions. Dask shows poor weak scaling due to the incompleteness of Dask with respect to NumPy operations. For example, assignments to Dask arrays are not supported by the library itself but are heavily utilized in the implemented LASSO algorithm. Consequently, Dask cannot make efficient use of its lazy evaluation concept~\cite{daskdocumentation} for this algorithm. The HeAT GPU version also does not scale well, albeit with a significantly lower runtime than Dask. This is due to the high number of communication operations required.

Strong scaling measurements (\cref{fig:scaling-lasso}, lower panel) were conducted for the entire sample set. The trends observed in the weak scaling measurements are also visible here. Dask shows almost no scaling, whereas the HeAT CPU measurements indicate a good scaling behaviour. For the  HeAT GPU implementation the speedup decreases with the increase in computing resources.  Overall, it is apparent that HeAT outperforms Dask by more than two orders of magnitude. 

\section{Discussion}
\label{sec:discussion}

We have presented HeAT, a Python-based framework for distributed data analytics and machine learning on multiple CPUs and GPUs. It offers transparent parallel data handling and operations to exploit the available hardware, be it personal workstations or supercomputers. The NumPy-like API enables users to easily translate existing NumPy code into distributed applications. As the user-base of HPC resources is predominately composed of domain experts with limited knowledge of parallel programming concepts, the number of exposed configuration parameters for parallel constructs is minimized. These constructs are both powerful and versatile enough for the implementation of a large variety of distributed algorithms.

PyTorch has been selected as an eager, node-local compute engine for HeAT. As a direct result, HeAT benefits from PyTorch's highly optimized functions. However, PyTorch does not offer parallel data decomposition or distributed algorithms. HeAT provides a custom communication layer for N-dimensional array objects and specifically designed hierarchical algorithms to leverage the full potential of PyTorch for distributed environments. This strategy enables the efficient utilization of the underlying hardware by exploiting locality in the memory hierarchy.

Memory distribution for scalable algorithms requires efficient communication between compute nodes at runtime via high-speed network links. HeAT is designed to leverage the potential of such systems. Significant efforts have been made to efficiently utilize the available hardware, including multi-GPU architectures, while avoiding central bottlenecks, such as workload schedulers, and excessive I/O, e.g. serial file access or journaling. Weak and strong scaling experiments on a number of applications (\cref{sec:performance-result}) demonstrate that HeAT consistently outperforms Dask by up to two orders of magnitude. Moreover, larger datasets caused Dask to raise memory errors in several experiments. HeAT enables users to access a substantial portion of maximum potential performance via a high-level interface, relatively independent of data characteristics. 

Coupling of Dask with CuPy, CuDF, or CuML can in principle be used for distributed GPU computation. Within the performance evaluations \cref{subsubsec:performance-mean}, a multi-GPU benchmark for Dask with CuPy was conducted for the mean calculation, which did not show any improvement. During these experiments it was found that this is only valid for a multi-node single-GPU setup. We hypothesize that Dask does not take care of copying the data from GPU to CPU before communicating and thus, the communication stack cannot properly access the VRAM. In order to enable multi-node multi-GPU support, data must be moved manually from CPU to GPU and back for every operation utilizing communication as demonstrated in \cref{lst:DaskCupy} (Appendix). For high-level algorithms, this is a very cumbersome task requiring a substantial understanding of distributed programming. 

As discussed previously, most frameworks for parallel deep learning applications primarily focus on data-parallelism. However, generalized model parallelism is not available to the authors' knowledge. HeAT's programming model facilitates straight-forward data-parallelism as well as model parallelism and pipelining. The use of a custom communication layer allows for the implementation of distributed automatic differentiation, which is a vital part for a distributed model architecture. First steps in this direction are already underway with \texttt{mpi4torch}~\cite{mpi4torch}, a prototype for distributed AD. 

This subsequently offers the opportunity for the development of other high-level differentiable algorithms. In light of the ever increasing need for machine learning models to yield reliable predictions, considerable efforts have been put towards the development of probabilistic approaches. HeAT's programming model and internal design give access to all levels of algorithmic development and by such offers an intuitive way to implement such approaches.

\section{Conclusion}
\label{sec:conclusion}

With HeAT, we address the needs of an ever-growing community of scientists, both in academia and industry, who seek to accelerate the process of extracting information from Big Data. To this end, we have set upon the task of combining data analytics and machine learning algorithms with state-of-the-art high-performance computing concepts into an easy-to-use Python library. 

We have demonstrated that even in its current early stage, HeAT offers great potential. The convergence of speed and usability sets it up to redefine high-performance data analytics by putting high levels of parallelism within reach of scientists in academia and industry alike. HeAT offers a way to easily develop application-specific algorithms while leveraging the available computational resources, setting it apart from other approaches and libraries.

\newpage
\section*{Acknowledgment}
\label{sec:acknowledgment}

The authors would like to thank the system administrators at the Jülich Supercomputing Centre and in particular Dr. Alexandre Strube for their continuous support in maintaining the benchmarking HPC system. Furthermore, we want to thank the Helmholtz Analytics Framework collaboration for thorough feedback and valuable suggestions.
 
\bibliographystyle{IEEEtranDOI}
\bibliography{main}

\begin{thebibliography}{10}
\providecommand{\url}[1]{#1}
\csname url@samestyle\endcsname
\providecommand{\newblock}{\relax}
\providecommand{\bibinfo}[2]{#2}
\providecommand{\BIBentrySTDinterwordspacing}{\spaceskip=0pt\relax}
\providecommand{\BIBentryALTinterwordstretchfactor}{4}
\providecommand{\BIBentryALTinterwordspacing}{\spaceskip=\fontdimen2\font plus
\BIBentryALTinterwordstretchfactor\fontdimen3\font minus
  \fontdimen4\font\relax}
\providecommand{\BIBforeignlanguage}[2]{{%
\expandafter\ifx\csname l@#1\endcsname\relax
\typeout{** WARNING: IEEEtran.bst: No hyphenation pattern has been}%
\typeout{** loaded for the language `#1'. Using the pattern for}%
\typeout{** the default language instead.}%
\else
\language=\csname l@#1\endcsname
\fi
#2}}
\providecommand{\BIBdecl}{\relax}
\BIBdecl

\bibitem{viranen2020scipy}
P.~{Virtanen}, R.~{Gommers}, T.~E. {Oliphant}, M.~{Haberland}, T.~{Reddy}
  \emph{et~al.}, ``{SciPy 1.0: Fundamental Algorithms for Scientific Computing
  in Python},'' \emph{Nature Methods}, vol.~17, pp. 261--272, 2020.

\bibitem{walt2011numpy}
S.~v.~d. Walt, S.~C. Colbert, and G.~Varoquaux, ``{The NumPy array: a Structure
  for Efficient Numerical Computation},'' \emph{Computing in Science \&
  Engineering}, vol.~13, no.~2, pp. 22--30, 2011.

\bibitem{abadi2015tensorflow}
\BIBentryALTinterwordspacing
M.~Abadi, A.~Agarwal, P.~Barham, E.~Brevdo, Z.~Chen \emph{et~al.},
  ``{TensorFlow: Large-Scale Machine Learning on Heterogeneous Systems},''
  2015, software available from tensorflow.org, [accessed at 2020-02-24].
  [Online]. Available: \url{http://tensorflow.org/}
\BIBentrySTDinterwordspacing

\bibitem{paszke2019pytorch}
\BIBentryALTinterwordspacing
A.~Paszke, S.~Gross, F.~Massa, A.~Lerer, J.~Bradbury \emph{et~al.}, ``{PyTorch:
  An Imperative Style, High-Performance Deep Learning Library},'' in
  \emph{{Advances in Neural Information Processing Systems 32}}.\hskip 1em plus
  0.5em minus 0.4em\relax Curran Associates, Inc., 2019, pp. 8024--8035.
  [Online]. Available:
  \url{http://papers.neurips.cc/paper/9015-pytorch-an-imperative-style-high-performance-deep-learning-library.pdf}
\BIBentrySTDinterwordspacing

\bibitem{mpi2015mpi}
\BIBentryALTinterwordspacing
{Message Passing Interface Forum}, \emph{MPI: A Message-Passing Interface
  Standard, Version 3.1}.\hskip 1em plus 0.5em minus 0.4em\relax High
  Performance Computing Center Stuttgart (HLRS), 2015. [Online]. Available:
  \url{https://fs.hlrs.de/projects/par/mpi//mpi31/}
\BIBentrySTDinterwordspacing

\bibitem{pedregosa2011sklearn}
F.~Pedregosa, G.~Varoquaux, A.~Gramfort, V.~Michel, B.~Thirion \emph{et~al.},
  ``{Scikit-learn: Machine Learning in Python},'' \emph{{Journal of Machine
  Learning Research}}, vol.~12, pp. 2825--2830, 2011.

\bibitem{lam2015numba}
S.~K. Lam, A.~Pitrou, and S.~Seibert, ``{Numba: A LLVM-based Python JIT
  Compiler},'' in \emph{{Proceedings of the Second Workshop on the LLVM
  Compiler Infrastructure in HPC}}, 2015, pp. 1--6.

\bibitem{nishino2017cupy}
R.~Nishino and S.~H.~C. Loomis, ``{CUPy: A NumPy-compatible Library for NVidia
  GPU Calculations},'' \emph{{Advances in Neural Information Processing Systems
  31}}, p. 151, 2017.

\bibitem{rapids}
\BIBentryALTinterwordspacing
R.~D. Team, \emph{{RAPIDS: Collection of Libraries for End to End GPU Data
  Science}}, 2018. [Online]. Available: \url{https://rapids.ai}
\BIBentrySTDinterwordspacing

\bibitem{reback2020pandas}
\BIBentryALTinterwordspacing
T.~P.~D. Team, ``{pandas-dev/pandas: Pandas},'' 2020, [accessed at 2020-03-17].
  [Online]. Available: \url{https://doi.org/10.5281/zenodo.3509134}
\BIBentrySTDinterwordspacing

\bibitem{chen2015mxnet}
\BIBentryALTinterwordspacing
T.~Chen, M.~Li, Y.~Li, M.~Lin, N.~Wang \emph{et~al.}, ``{MXNet: A Flexible and
  Efficient Machine Learning Library for Heterogeneous Distributed Systems},''
  2015, [accessed at 2020-02-02]. [Online]. Available:
  \url{http://arxiv.org/abs/1512.01274}
\BIBentrySTDinterwordspacing

\bibitem{bradburg2018jax}
\BIBentryALTinterwordspacing
J.~Bradbury, R.~Frostig, P.~Hawkins, M.~J. Johnson, C.~Leary \emph{et~al.},
  ``{JAX: Composable Transformations of Python+NumPy Programs},'' 2018,
  [accessed at 2020-02-29]. [Online]. Available:
  \url{http://github.com/google/jax}
\BIBentrySTDinterwordspacing

\bibitem{rajbh2019zero}
\BIBentryALTinterwordspacing
S.~Rajbhandari, J.~Rasley, O.~Ruwase, and Y.~He, ``{ZeRO: Memory Optimization
  Towards Training A Trillion Parameter Models},'' 2019, [accessed at
  2020-03-17]. [Online]. Available: \url{https://arxiv.org/abs/1910.02054}
\BIBentrySTDinterwordspacing

\bibitem{sergeev2018horovod}
\BIBentryALTinterwordspacing
A.~Sergeev and M.~Del~Balso, ``{Horovod: Fast and Easy Distributed Deep
  Learning in TensorFlow},'' 2018, [accessed at 2020-03-17]. [Online].
  Available: \url{http://arxiv.org/abs/1802.05799}
\BIBentrySTDinterwordspacing

\bibitem{tohid2018phylanx}
R.~Tohid, B.~Wagle, S.~Shirzad, P.~Diehl, A.~Serio \emph{et~al.},
  ``{Asynchronous Execution of Python Code on Task-Based Runtime Systems},'' in
  \emph{{2018 IEEE/ACM 4th International Workshop on Extreme Scale Programming
  Models and Middleware (ESPM2)}}.\hskip 1em plus 0.5em minus 0.4em\relax IEEE,
  2018, pp. 37--45.

\bibitem{chen2017benchmarking}
L.~Chen, B.~Peng, B.~Zhang, T.~Liu, Y.~Zou \emph{et~al.}, ``{Benchmarking
  Harp-DAAL: High Performance Hadoop on KNL clusters},'' in \emph{2017 IEEE
  10th International Conference on Cloud Computing (CLOUD)}.\hskip 1em plus
  0.5em minus 0.4em\relax IEEE, 2017. doi:
  https://doi.org/10.1109/CLOUD.2017.19 pp. 82--89.

\bibitem{dean2008mapreduce}
J.~Dean and S.~Ghemawat, ``{MapReduce: Simplified Data Processing on Large
  Clusters},'' \emph{{Communications of the ACM}}, vol.~51, no.~1, pp.
  107--113, 2008.

\bibitem{bauer2019legate}
M.~Bauer and M.~Garland, ``{Legate NumPy: Accelerated and Distributed Array
  Computing},'' in \emph{Proceedings of the International Conference for High
  Performance Computing, Networking, Storage and Analysis}, 2019. doi:
  https://doi.org/10.1145/3295500.3356175 pp. 1--23.

\bibitem{rocklin2015dask}
M.~Rocklin, ``{Dask: Parallel Computation with Blocked algorithms and Task
  Scheduling},'' in \emph{Proceedings of the 14th Python in Science Conference
  (SciPy 2015)}, K.~Huff and J.~Bergstra, Eds., 2015, pp. 130--136.

\bibitem{spark}
M.~Zaharia, R.~Xin, P.~Wendell, T.~Das, M.~Armbrust \emph{et~al.}, ``{Apache
  Spark: A Unified Engine for Big Data Processing},'' \emph{Communications of
  the ACM}, vol.~59, no.~11, p. 56–65, 2016.

\bibitem{sparkvdask}
\BIBentryALTinterwordspacing
M.~Dugré, V.~Hayot-Sasson, and T.~Glatard, ``A performance comparison of dask
  and apache spark for data-intensive neuroimaging pipelines,'' 2019, [accessed
  at 2020-03-17]. [Online]. Available: \url{https://arxiv.org/abs/1907.13030}
\BIBentrySTDinterwordspacing

\bibitem{darema2001spmd}
F.~Darema, ``{The SPMD Model: Past, Present and Future},'' in \emph{Recent
  Advances in Parallel Virtual Machine and Message Passing Interface}, ser.
  Lecture Notes in Computer Science, Y.~Cotronis and J.~Dongarra, Eds., vol.
  2131, no.~1.\hskip 1em plus 0.5em minus 0.4em\relax Springer Berlin
  Heidelberg, 2001. doi: https://doi.org/10.1007/3-540-45417-9\_1. ISBN
  978-3-540-45417-5 pp. 1--1.

\bibitem{valiant1990bsp}
L.~Valiant, ``{A Bridging Model for Parallel Computation},''
  \emph{Communications of the ACM}, vol.~33, no.~8, pp. 103--111, 1990.

\bibitem{dagum1998openmp}
L.~Dagum and R.~Menon, ``{OpenMP: an Industry Standard API for Shared-memory
  Programming},'' \emph{{IEEE Computational Science and Engineering}}, vol.~5,
  no.~1, pp. 46--55, 1998.

\bibitem{pheatt2008tbb}
C.~Pheatt, ``{Intel{\textregistered} Threading Building Blocks},''
  \emph{{Journal of Computing Sciences in Colleges}}, vol.~23, no.~4, pp.
  298--298, 2008.

\bibitem{nickolls2008cuda}
J.~Nickolls, I.~Buck, M.~Garland, and K.~Skadron, ``{Scalable Parallel
  Programming with CUDA},'' \emph{Queue}, vol.~6, no.~2, pp. 40--53, 2008.

\bibitem{dalcin08mpi4py}
\BIBentryALTinterwordspacing
L.~Dalc\'{i}n, R.~Paz, M.~Storti, and J.~D'El\'{i}a, ``{MPI for Python:
  Performance Improvements and MPI-2 Extensions},'' \emph{{Journal of Parallel
  and Distributed Computing}}, vol.~68, pp. 655--662, 05 2008. [Online].
  Available: \url{https://doi.org/10.1016/j.jpdc.2007.09.005}
\BIBentrySTDinterwordspacing

\bibitem{cordts2016cityscapes}
M.~Cordts, M.~Omran, S.~Ramos, T.~Rehfeld, M.~Enzweiler \emph{et~al.}, ``{The
  Cityscapes Dataset for Semantic Urban Scene Understanding},'' in \emph{{2016
  IEEE Conference on Computer Vision and Pattern Recognition (CVPR)}}.\hskip
  1em plus 0.5em minus 0.4em\relax IEEE, June 2016. doi: 10.1109/CVPR.2016.350.
  ISSN 1063-6919 pp. 3213--3223.

\bibitem{baldi2014searching}
P.~Baldi, P.~Sadowski, and D.~Whiteson, ``{Searching for Exotic Particles in
  High-energy Physics with Deep Learning},'' \emph{{Nature Communications}},
  vol.~5, p. 4308, 2014.

\bibitem{uciMachineLearningRepository}
\BIBentryALTinterwordspacing
D.~Dua and C.~Graff, ``{UCI Machine Learning Repository},'' 2017, [accessed at
  2020-03-17]. [Online]. Available:
  \url{http://archive.ics.uci.edu/ml/datasets/SUSY}
\BIBentrySTDinterwordspacing

\bibitem{elbern20004dvar}
H.~Elbern, H.~Schmidt, O.~Talagrand, and A.~Ebel, ``{4D-variational Data
  Assimilation with an Adjoint Air Quality Model for Emission Analysis},''
  \emph{Environmental Modelling \& Software}, vol.~15, no.~6, pp. 539--548,
  2000.

\bibitem{hdf52020hdf5}
\BIBentryALTinterwordspacing
{The HDF Group}, ``{Hierarchical Data Format, version 5},'' 1997, [accessed at
  2020-02-29]. [Online]. Available: \url{http://www.hdfgroup.org/HDF5/}
\BIBentrySTDinterwordspacing

\bibitem{collette2014h5py}
A.~Collette, \emph{{Python and HDF5}}.\hskip 1em plus 0.5em minus 0.4em\relax
  O'Reilly, 2013. ISBN 978-1449367831

\bibitem{daskdocumentation}
\BIBentryALTinterwordspacing
{Dask Development Team}, ``{Dask Documentation},'' 2020, [accessed at
  2020-08-07]. [Online]. Available: \url{https://docs.dask.org/en/latest/}
\BIBentrySTDinterwordspacing

\bibitem{moments}
J.~Bennett, R.~W. Grout, P.~P. P{\'e}bay, D.~C. Roe, and D.~C. Thompson,
  ``{Numerically Stable, Single-pass, Parallel Statistics Algorithms},''
  \emph{{2009 IEEE International Conference on Cluster Computing and
  Workshops}}, pp. 1--8, 2009.

\bibitem{debus2020high}
C.~Debus, A.~Ruettgers, A.~Petrarolo, M.~Kobald, and M.~Siggel,
  ``{High-performance Data Analytics of Hybrid Rocket Fuel Combustion Data
  Using Different Machine Learning Approaches},'' in \emph{{AIAA Scitech 2020
  Forum}}, 2020. doi: https://doi.org/10.2514/6.2020-1161 p. 1161.

\bibitem{macqueen1967kmeans}
J.~MacQueen, ``{Some Methods for Classification and Analysis of Multivariate
  Observations},'' in \emph{{Proceedings of the Fifth Berkeley Symposium on
  Mathematical Statistics and Probability}}, vol.~1, no.~14, 1967, pp.
  281--297.

\bibitem{lloyd1982least}
S.~Lloyd, ``{Least Squares Quantization in PCM},'' \emph{IEEE Transactions on
  Information Theory}, vol.~28, no.~2, pp. 129--137, 1982.

\bibitem{mpi4torch}
\BIBentryALTinterwordspacing
{Philipp Knechtges}, ``{MPI4Torch: an automatic-differentiable wrapper of MPI
  functions for the PyTorch tensor library},'' 2020, [accessed at 2020-08-11].
  [Online]. Available: \url{https://github.com/helmholtz-analytics/mpi4torch}
\BIBentrySTDinterwordspacing

\end{thebibliography}
\section*{Appendix}
\label{sec:appendix}

\begin{lstlisting}[
    caption={Code for multi-node multi-GPU calculation of the mean with Dask with CuPy coupling.},
    label={lst:DaskCupy}
]
import cupy
import dask.array as da
import h5py

from dask.distributed import Client
from mpi4py import MPI

NUM_GPUS = 4

def gpu_part(x, items):
    rank = MPI.COMM_WORLD.rank
    with cupy.cuda.Device(rank % NUM_GPUS):
        x_gpu = cupy.asarray(x)
        return (x_gpu / items).sum(axis=0)

def cpu_part(x):
    return cupy.asnumpy(x)

# connect dask client
client = Client(scheduler_file="scheduler.json")

# load data
handle = h5py.File("cityscapes_300.h5", "r")
data = da.from_array(
            handle["cityscapes_data"], 
            chunks=('auto', -1,)
       ).persist()

# calculate mean
mean = data.map_blocks(gpu_part, data.shape[0])
mean = mean.map_blocks(cpu_part).sum(axis=0)
mean.compute()
\end{lstlisting}

\begin{lstlisting}[
    caption={Code for multi-node multi-GPU calculation of the mean with HeAT.},
    label={lst:HeatMean}
]
import heat as ht
import h5py
import torch

# select GPU if available
if torch.cuda.device_count() > 0:
    ht.use_device("gpu")

# load data
split_axis = 0
data = ht.load(
    "cityscapes_300.h5", 
    split=split_axis,
    dataset="cityscapes_data",
)

# calculate mean
mean = data.mean()
\end{lstlisting}

\end{document}